\pdfoutput=1

\documentclass[11pt,twoside,a4paper,cmspaper,final,collab]{cms-tdr}
\def\svnVersion{60d0f14-D}\def\svnDate{2020/11/05}\def\cmsCernNoTag{CERN-EP-2019-261}\def\cmsCernDate{\today}\def\cmsMessage{"Published in Computing and Software for Big Science as \href{http://dx.doi.org/10.1007/s41781-020-00041-z}{\doi{10.1007/s41781-020-00041-z}.}"}

\begin{document}\cmsNoteHeader{HIG-18-027}

\hyphenation{had-ron-i-za-tion}
\hyphenation{cal-or-i-me-ter}
\hyphenation{de-vices}
\ifthenelse{\boolean{cms@external}}{\providecommand{\cmsLeft}{upper\xspace}}{\providecommand{\cmsLeft}{left\xspace}}
\ifthenelse{\boolean{cms@external}}{\providecommand{\cmsRight}{lower\xspace}}{\providecommand{\cmsRight}{right\xspace}}

\providecommand{\pz}{\ensuremath{p_z}\xspace}
\providecommand{\pTjreco}{\ensuremath{\pt^\text{reco}}}
\providecommand{\pTjgen}{\ensuremath{\pt^{\text{gen}}}}
\providecommand{\abs}[1]{\ensuremath{|#1|}}
\providecommand{\E}{\ensuremath{E}}

\providecommand{\ZHbbll}{\ensuremath{\PZ(\to \ell^+\ell^-)\PH(\to \PQb\PAQb)}\xspace}
\providecommand{\HHbbgg}{\ensuremath{\PH(\to\PQb\PAQb)\PH(\to\PGg\PGg)}\xspace}

\newcommand{\sbar}{\ensuremath{\overline{\mathrm{s}}}\xspace}

\cmsNoteHeader{HIG-18-027}
\title{A deep neural network for simultaneous estimation of \PQb jet energy and resolution}
\titlerunning{Neural net estimation of \PQb energy and resolution}

\date{\today}

\abstract{
We describe a method to obtain point and dispersion estimates for the energies of jets arising from \PQb quarks produced in proton-proton collisions at an energy of $\sqrt{s}=13\TeV$ at the CERN LHC. The algorithm is trained on a large sample of simulated \PQb jets and validated on data recorded by the CMS detector in 2017 corresponding to an integrated luminosity of 41\fbinv. A multivariate regression algorithm based on a deep feed-forward neural network employs jet composition and shape information, and the properties of reconstructed secondary vertices associated with the jet. The results of the algorithm are used to improve the sensitivity of analyses that make use of \PQb jets in the final state, such as the observation of Higgs boson decay to $\bbbar$.
}

\hypersetup{
pdfauthor={CMS Collaboration},
pdftitle={A deep neural network for simultaneous estimation of b jet energy and resolution},
pdfsubject={CMS},
pdfkeywords={CMS, b jets, Higgs boson, jet energy, jet resolution, deep learning}}
\maketitle

\section{Introduction}
\label{sec:introduction}

Following the discovery of the 125\GeV Higgs boson reported by the ATLAS and CMS Collaborations at the CERN LHC in 2012~\cite{Aad:2012tfa, Chatrchyan:2012xdj,Chatrchyan:1529911}, a rich research program was established to probe this new particle. The program includes the measurement of all production and decay modes that are accessible at the LHC. The decay of the Higgs boson into a pair of vector bosons was established with a statistical significance higher than five standard deviations individually for photon, {\PZ} and {\PW} pairs using data collected at the LHC from 2011 to 2013 at center-of-mass energies of $\sqrt{s}=7$ and 8\TeV~\cite{Aad:2014eva,ATLAS:2014aga,Aad:2014eha,Chatrchyan:2013mxa,Chatrchyan:2013iaa,Khachatryan:2014ira}. A few years later, the combination of CMS data sets collected at  8 and 13\TeV was used to report the observation of Higgs boson decay to a pair of $\tau$ leptons~\cite{Sirunyan:2017khh}, followed by the observation of the associated production of a Higgs boson with a top quark-antiquark pair (\ttbar)~\cite{Aaboud:2018urx,Sirunyan:2018hoz}.

Higgs boson decay to a \PQb quark-antiquark pair ($\bbbar$) was only recently announced by the CMS~\cite{Sirunyan:2018kst} and ATLAS~\cite{Aaboud:2018zhk} Collaborations, despite it being the dominant decay mode. This is because of the challenges associated with separating the signal from the large background of $\bbbar$ produced by quantum chromodynamics (QCD) processes. Good resolution of the reconstructed invariant mass of Higgs boson candidates is necessary to have a more favorable signal-to-background ratio.
This is achieved in CMS by the method described in this paper, based on a deep neural network (DNN) that estimates the energy of jets originating from \PQb quarks (\PQb jets). Similar algorithms, using neural networks, were previously used by the CDF Collaboration at the Tevatron~\cite{Aaltonen:2011bp0,Aaltonen:2011bp}, and BDT-based energy regressions were used earlier by the CMS Collaboration to estimate the energy of \PQb jets~\cite{Khachatryan:2015bnx}.

The approach described in this paper is to use a regression algorithm that is implemented in a feed-forward neural network with six hidden layers trained on a very large
data set, consisting of Monte Carlo (MC) simulated \PQb jets. The algorithm has a considerably larger modeling capability than those used previously. This approach was made possible by leveraging recent advances in hardware accelerators, such as graphics processing units (GPU), and in modern packages for automatic differentiation to handle the otherwise expensive computations involved in this task. A minimization of a loss function that combines a Huber~\cite{Huber:1964} and two quantile~\cite{RePEc} loss terms, enables simultaneous training of point and dispersion estimators of the regression target without making any assumptions about the functional form of its distribution. The point estimator is used as a correction of the measured \PQb jet energy while dispersion estimators are used to build a jet-by-jet resolution estimate. The CMS collaboration had previously developed a BDT-based approach to estimate the energy and per-object resolution~\cite{Khachatryan:2015iwa,Khachatryan:2015hwa,Sirunyan:2019kia}. This can be achieved by training separate regressions to obtain energy and per-object resolution estimators, or by means of a semiparametric regression~\cite{Khachatryan:2015iwa,Khachatryan:2015hwa}. For a semiparametric regression, the training relies on the knowledge of the analytical shape of the target distribution. The novel characteristic of the algorithm described in this paper is the simultaneous training of the point and dispersion estimators without reference to an ansatz distribution for the regression target. This method is validated on data collected by the CMS detector in 2017.

In the following, Section~\ref{sec:cms_detector} and Section~\ref{sec:cms_datasets} describe the CMS detector and the data sets used for this work. The regression problem and the inputs are described in Section~\ref{sec:inputs_and_prob}. In Section~\ref{sec:learning_setup} the loss function is introduced, while the DNN architecture and its training are summarized in Section~\ref{sec:NN_arc}. Finally, the results are presented in Section~\ref{sec:Results}, followed by the summary in Section~\ref{sec:Conclusion}.

\section{The CMS detector}
\label{sec:cms_detector}

The central feature of the CMS detector is a superconducting solenoid of 6\unit{m} internal diameter,
providing a magnetic field of 3.8\unit{T}. Within the solenoid volume are a silicon pixel and strip
tracker, a lead tungstate crystal electromagnetic calorimeter (ECAL), and a brass and
scintillator hadron calorimeter (HCAL), each composed of a barrel and two endcap
sections. Forward calorimeters extend the pseudorapidity ($\eta$) coverage provided by the barrel
and endcap detectors. Muons are detected in gas-ionization chambers embedded in the steel
flux-return yoke outside the solenoid.
A detailed description of the apparatus, together with a definition of the coordinate system used and the
relevant kinematic variables, can be found in Ref.~\cite{Chatrchyan:2008aa}.

The particle-flow  (PF) algorithm~\cite{Sirunyan:2017ulk} used by CMS aims to reconstruct and identify each
individual particle in an event, with an optimized combination of information from the
various elements of the CMS detector. Photon energies are obtained from ECAL
data.
The candidate vertex with the largest value of summed physics-object $\pt^2$ is taken to be the primary proton-proton ($\Pp\Pp$) interaction vertex.
The energy of each electron in the event is determined from a combination of the electron
momentum at the primary interaction vertex,
as determined by the tracker, the energy of the
corresponding ECAL cluster, and the energy sum of all bremsstrahlung photons spatially
compatible with having originated from the electron. The momentum of each muon is obtained via
the curvature of the corresponding track. The energy of each charged hadron is determined from
a combination of momentum measured in the tracker and the matching ECAL and HCAL
energy deposits, corrected for zero-suppression effects and for the response function of
the calorimeters to hadronic showers. Finally, for a neutral hadron the energy is obtained
from the corresponding  HCAL corrected energies.
The anti-\kt algorithm~\cite{Cacciari:2008gp,Cacciari:2011ma} with a distance parameter of 0.4 is applied offline to the full set of PF candidates in order to cluster them into jets.  The jet momentum is determined by the vectorial sum of all particle momenta in the jet. The jet energy resolution  typically amounts to 15--20\% at 30\GeV, 10\% at 100\GeV, and 5\% at 1\TeV~\cite{Khachatryan_2017}.

Additional $\Pp\Pp$ interactions within the same
or nearby bunch crossings (pileup)
can contribute unrelated particles to the jet. To mitigate the effects of pileup, charged particles with tracks originating from pileup vertices are discarded before jet reconstruction. Then, the residual contamination from neutral particles and charged particles without reconstructed tracks is estimated for each event and subtracted from the jet energy.
Jet energy corrections are derived from simulation to
bring the measured average response for jets in line with particle-level jets. Neutrinos are not included in the clustering of particle-level jets. In-situ
measurements of the transverse momentum balance in dijet, photon+jet, $\PZ$+jet, and multijet events
are used to account for residual differences between the jet energy scales in data and
simulation~\cite{Chatrchyan:2011ds}. We refer to this correction algorithm as the
baseline algorithm.

\section{Data sets}
\label{sec:cms_datasets}

The DNN was trained on 100 million \PQb jets from a simulated sample of \ttbar events produced in {\Pp}{\Pp} collisions at $\sqrt{s}=13\TeV$, generated at next-to-leading-order (NLO) accuracy in perturbative QCD (pQCD) with the
\POWHEG~v2 program~\cite{Campbell:2014kua}. Predictions of the model were then tested on simulated events with \PQb jets coming from a variety of physical processes in order to validate performance
in all relevant kinematic regions. To this end, \PQb jets from the decay of Higgs bosons produced in association with a {\PZ} boson, \ZHbbll, where $\ell$ is an electron or a muon, were generated with
the \MGvATNLO generator~\cite{Alwall:2014hca} at NLO pQCD accuracy. Additionally, \PQb jets from the decay of Higgs boson pairs produced either from gluon fusion or in the decay of a new, spin 0 resonance, with one Higgs boson decaying to a \PQb quark-antiquark pair and the other to a pair of photons, \HHbbgg, were generated with \MGvATNLO at leading-order accuracy in pQCD.

Two definitions of jets are used in this study: ``generator-level jets'', clustered from stable
particles produced by the MC generator that include the contribution from the neutrino's momentum, and  ``reconstructed jets'', clustered from reconstructed
particle-flow candidates. The reconstructed \PQb jets were matched to generated \PQb jets to avoid contamination by light flavoured jets. 
For each reconstructed jet the corresponding generator-level jet is found by spatial matching in the $\eta-\phi$ plane by requiring the distance $\Delta R = \sqrt{\smash[b]{(\Delta\eta)^2+(\Delta\phi)^2}}$ (where $\phi$ is the azimuthal angle in radians) to be $\Delta R < 0.4$.  
The reconstructed \PQb jets were then selected by applying a minimum threshold for transverse
momentum ($\pTjreco > 15$\GeV, $\pTjgen > 15$\GeV) and by requiring the pseudorapidity of the central axis of the reconstructed jet to be within the tracker acceptance ($\abs{\eta} < 2.4$).

Finally, to validate the regression model on data,
the output of the DNN for simulated \PQb jets was compared to that obtained for \PQb jets recorded by the CMS detector. The events used for this validation were recorded in 2017 with triggers~\cite{Khachatryan_trig} that require  the presence of at least one lepton.
This data set, corresponding to an integrated luminosity of 41\fbinv, was further enriched in {\PZ} bosons produced in association with \PQb jets.
The corresponding simulated events come from a sample of {\PZ} bosons and up to two additional partons generated with \MGvATNLO at NLO accuracy in pQCD.

For all simulated events, \PYTHIA~8.2~\cite{Sjostrand:2014zea} with the CP5
tune~\cite{Khachatryan:2015pea} is used for parton showering and hadronization. The CMS detector
response is simulated by the \GEANTfour~\cite{Agostinelli:2002hh} package, and simulated
pileup interactions are added to the hard-scattering process to match the distribution
of pileup interactions observed in data, for which the observed mean number of interactions per bunch crossing is 32.

\section{Energy regression and input features}
\label{sec:inputs_and_prob}

In comparison to jets arising from light-flavor quarks or gluons, jets arising from \PQb quarks have special characteristics that call for dedicated energy corrections. In particular, \PQb jets contain \PQb hadrons that can often decay to a final state with a charged lepton and a neutrino. The neutrinos, which only interact via the weak force, escape detection, leading to an underestimate of the \PQb jet energy, with a corresponding  degradation of energy resolution. As described in Section~\ref{sec:cms_detector}, the jet energy is reconstructed by clustering its  constituents within a given distance parameter. Compared to jets originating from light-flavor quarks and gluons, \PQb jets, because of their higher mass, tend to spread radially over a wider area in the $\eta$-$\phi$ plane. This often leads to leakage of energy outside of the jet clustering region, further impacting the jet energy response and resolution.

\begin{figure}[thb]
\centering
\includegraphics[width=0.49\textwidth]{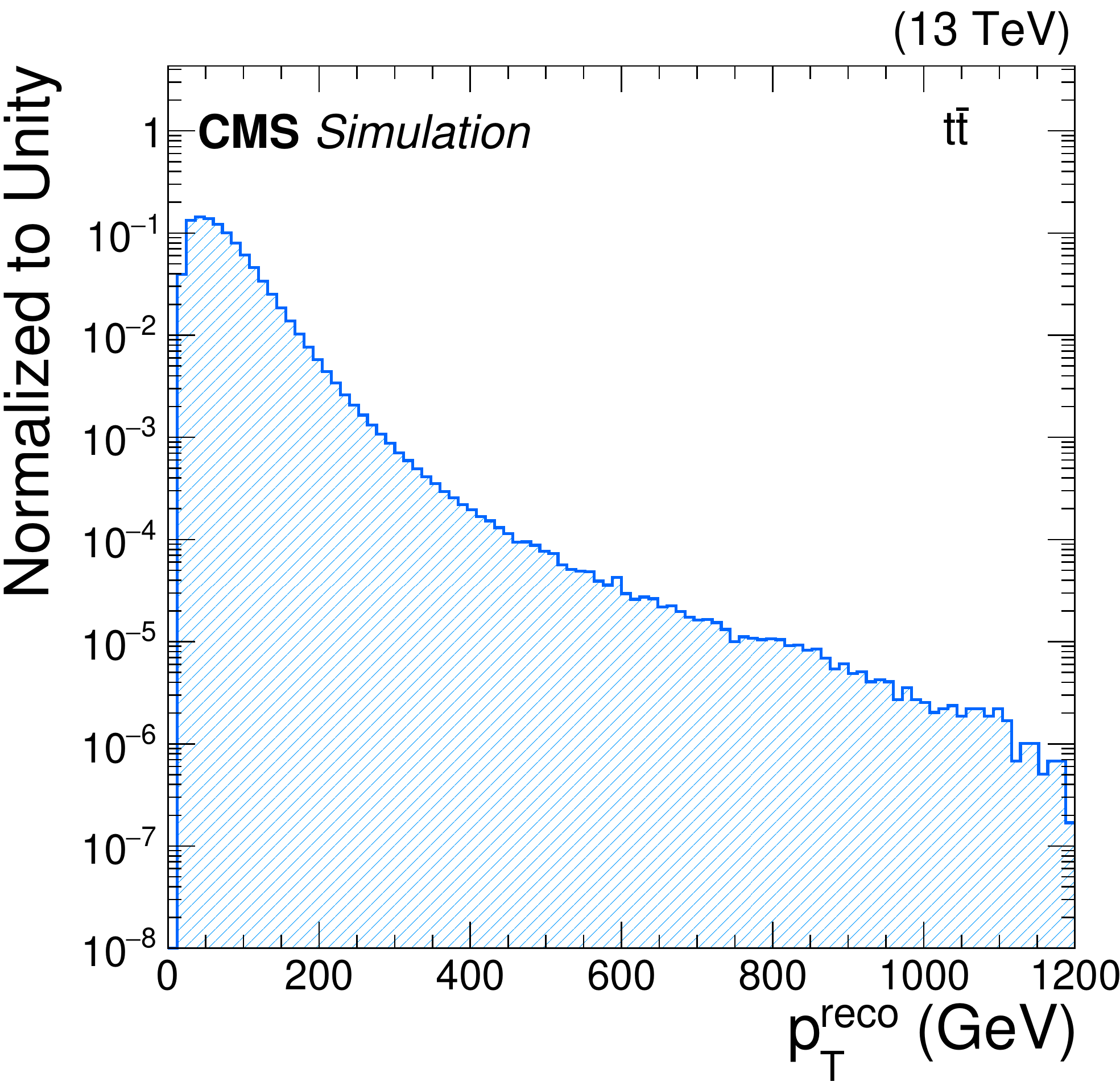}
\includegraphics[width=0.49\textwidth]{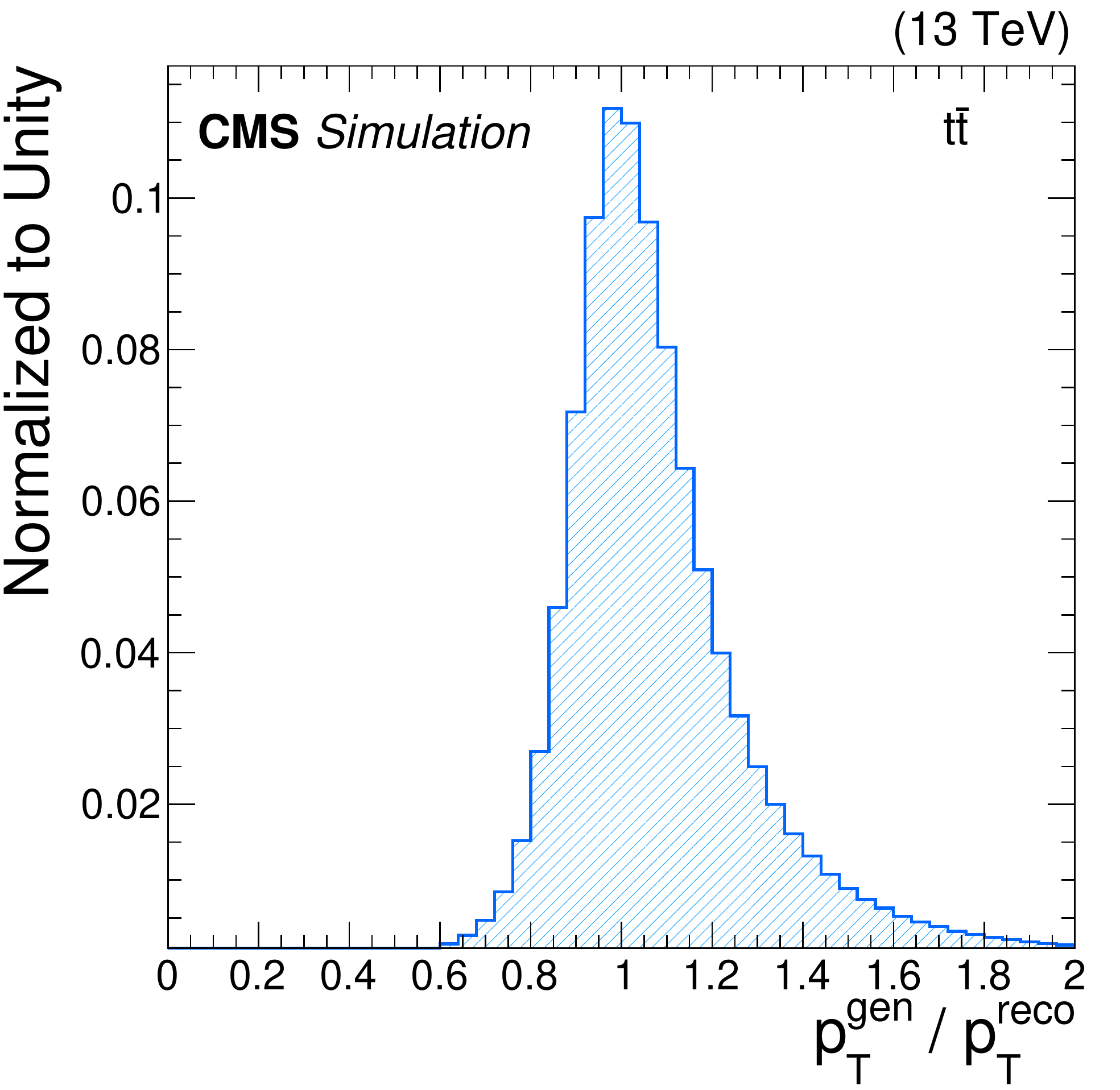}
\caption{(\cmsLeft) The $\pTjreco$ distribution for reconstructed \PQb jets in an MC \ttbar sample. (\cmsRight) Distribution of the regression target for the MC \ttbar training sample. \label{fig:bjets} }
\end{figure}

The \PQb jets used for the DNN  training come from a sample of simulated top quark events. The top quark decays before hadronising with a branching fraction close to unity into a \PQb jet and a {\PW} boson. At  LHC energies, it provides a source of \PQb jets that spans a large transverse momentum ($\pt$) spectrum and covers the full $\eta$ acceptance of the detector. The $\pTjreco$ value is corrected with the baseline algorithm as described in Section~\ref{sec:cms_detector}. Figure~\ref{fig:bjets} (\cmsLeft) shows the distribution of  $\pTjreco$,  for the selected \PQb jets.

The regression target, $y$, used in this study is defined as the ratio of the transverse momentum of the generator-level jet, $\pTjgen$, to that of the reconstructed jet, $\pTjreco$, applying the baseline jet energy corrections. Using this definition rather than using $\pTjgen$ directly has the effect of greatly reducing the variance of the target while producing a numerical value of order 1. The distribution of the target for \PQb jets from an MC simulated \ttbar sample is shown in Fig.~\ref{fig:bjets} (\cmsRight). To improve the convergence of the training of the DNN, the target is further standardized by subtracting its median value and dividing it by its standard deviation.

The DNN training inputs provide information about the kinematics, shape and composition of reconstructed jets. The inputs consist of the following features:
\begin{itemize}
\item jet kinematics: jet \pt, $\eta$, mass, and transverse mass, defined as $\sqrt{\smash[b]{ E^2 - \pz^2} }$;
\item information about pileup interactions: the median energy density in
    the event, $\rho$, corresponding to the amount of transverse momentum  per unit area that is due to overlapping collisions~\cite{Cacciari:2007fd};
\item information about semileptonic decays of \PQb  hadrons when an electron or muon
    candidate is clustered within a jet: the transverse component of lepton momentum
    perpendicular to the jet axis, the distance $\Delta R = \sqrt{\smash[b]{(\Delta\eta)^2+(\Delta\phi)^2}}$,
and a categorical variable that encodes information about the lepton candidate's flavour;
\item information about the secondary vertex, selected as the highest \pt displaced vertex linked to the jet: number of tracks associated to the vertex, transverse momentum and mass (computed assigning the pion mass to all reconstructed tracks forming the secondary vertex); the distance between the collision vertex and the secondary vertex computed in three dimensional space with its associated uncertainty~\cite{Chatrchyan:1704291,Sirunyan:2017ezt};
\item jet composition: largest $\pt$ value of any charged hadron candidates, fractions of energy carried by jet constituents; namely charged hadrons, neutral hadrons, muons, and an electromagnetic component coming from electrons and photons. These fractions are computed for the whole jet, and separately in five rings of $\Delta R$ around the jet axis ($\Delta R = $ 0--0.05, 0.05--0.1, 0.1--0.2, 0.2--0.3, 0.3--0.4);
\item multiplicity of PF candidates clustered to form the jet;
\item information about jet energy sharing among the jet constituents computed as
\begin{linenomath*}
\begin{equation}
\frac{\sqrt{\sum_ip_{T,i}^2}}{{\sum_ip_{T,i}}},
\end{equation}
\end{linenomath*}
where $i$ runs over all jet constituents.
\end{itemize}
This results in a total of 41 input features. No additional preprocessing is performed, apart from the input normalization provided by batch normalization~\cite{Ioffe:2015ovl} at the input layer of the DNN.

\section{Loss function}
\label{sec:learning_setup}
A possible approach to such a regression problem is to develop separate dedicated regressions to obtain energy and per-object resolution estimators. If the target distribution can be parametrized analytically, one can use a semiparametric regression to obtain estimates of the function parameters. This method has been used by the CMS collaboration to estimate the energy and resolution of electron and photon candidates~\cite{Khachatryan:2015iwa,Khachatryan:2015hwa}. Whereas for the photon and electron candidates the energy response can be parametrized by an analytically integrable function, this is less straightforward for \PQb jets, making such an approach to the problem more expensive computationally. An alternative approach is to simultaneously obtain point and dispersion estimates of the \PQb jet energy by defining a loss function that is completely agnostic to the target distribution. The correction to be applied to the reconstructed \PQb jet energy can be obtained as the estimated mean, while the per-jet \PQb jet energy resolution can be estimated as half the difference of the 75 and 25\% quantiles. Therefore, the regression loss function should provide the mean estimator ($\hat{y}$) and the 25 and 75\% quantiles of the target distribution.

The Huber loss function is employed to learn the mean of the target distribution via a minimization process. It is preferable to the mean squared error because of
its reduced sensitivity to the tails of the target distribution. It is defined as:
\begin{linenomath*}
\begin{equation}
\centering
H_{\delta}(z) = \begin{cases}
\frac{1}{2} z ^2, &\text{if $\abs{z} < \delta$;}\\
\delta\abs{z} - \frac{1}{2}\delta^2, &\text{otherwise,}
\end{cases}
\label{eq:Huber}
\end{equation}
\end{linenomath*}
where $z = y -\hat{y}$, and $\delta$ is set to 1 in our case.
To estimate the 25 and 75\% quantiles of the target distribution, the quantile loss function is used:
\begin{linenomath*}
\begin{equation}
\centering
\rho_{\tau}(z) = \begin{cases}
\tau  z, &\text{if $z > 0$;}\\
(\tau - 1) z, &\text{otherwise,}
\end{cases}
\label{eq:Quantile}
\end{equation}
\end{linenomath*}
where $\tau$ = 0.25 (0.75) corresponds to the 25 (75)\% quantile.

The complete loss function can then be written as:
\ifthenelse{\boolean{cms@external}}{
\begin{multline}
\text{loss}(\hat{y},\hat{y}_{25\%},\hat{y}_{75\%}) =
 \E_{(x,y) \sim p(x,y)}[H_{1}( y - \hat{y}(x) )\\+\rho_{0.25}( y - \hat{y}_{25\%}(x))+\rho_{0.75}( y -\hat{y}_{75\%}(x))],
\label{eq:Loss}
\end{multline}
}
{
\begin{linenomath*}
\begin{equation}
\text{loss}(\hat{y},\hat{y}_{25\%},\hat{y}_{75\%}) = \E_{(x,y) \sim p(x,y)} [H_{1}( y - \hat{y}(x) )+\rho_{0.25}( y - \hat{y}_{25\%}(x))+\rho_{0.75}( y -\hat{y}_{75\%}(x))],
\label{eq:Loss}
\end{equation}
\end{linenomath*}
}
where $\E_{(x,y) \sim p(x,y)}$ denotes the expectation value when sampling $(x,y)$ on the distribution $p(x,y)$, $x$ denotes the set of input features, and $p(x,y)$ is the joint distribution of the
input features and the target variables $y$ in the training sample. The symbols
$\hat{y}(x)$, $\hat{y}_{25\%}(x)$ and $\hat{y}_{75\%}(x)$ denote the DNN outputs: $\hat{y}(x)$ is the mean estimator, $\hat{y}_{25\%}(x)$ and $\hat{y}_{75\%}(x)$ are the 25 and 75\% quantile estimators, respectively.

\section{Neural network architecture}
\label{sec:NN_arc}

The model used for this study is a feed-forward, fully connected DNN with 6 hidden layers, 41 input features and 3 outputs: the energy correction and the 25 and 75\% quantiles.
As mentioned above, a batch normalization layer is applied at the DNN input.

Each hidden layer of the DNN is built from the following components:
\begin{itemize}
\item Dense layer: defined as a linear combination of all outputs from the previous layer.
\item Batch normalization layer: to transform the inputs to zero-mean and unit-variance.
\item Dropout unit: an operation that zeroes a fixed fraction of randomly chosen nodes during the training, used as a regularization handle. The dropout rate is one of the optimized hyperparameters of the DNN.
\item Activation unit: we chose the ``Leaky'' Rectified Linear
    Unit~(LReLU)~\cite{Maas:2013}:
\begin{linenomath*}
    \begin{equation}
        \centering
        \textrm{LReLU}(x) = \begin{cases}
            x, &\text{if $x \ge 0$};\\
            \beta x, &\text{if $x < 0$},
        \end{cases}
        \label{eq:ReLU}
    \end{equation}
\end{linenomath*}
    with $\beta = 0.2$.
\end{itemize}
A small slope $\beta$ = 0.2 was chosen for the LReLU to allow for a nonvanishing gradient over the domain of the function~\cite{Maas:2013}. The output layer has a linear activation function. The DNN is implemented using the \textsc{Keras} package~\cite{chollet2015keras} with \textsc{TensorFlow}
backend~\cite{tensorflow2015-whitepaper}. Back-propagation is done using stochastic gradient descent with the Adam
optimizer~\cite{Adam}.

\subsection{Hyperparameter optimization\label{sec:Optimization}}

To optimize the performance of the DNN, three hyperparameters
are considered: the depth of the network architecture, the dropout rate, and the gradient descent learning rate.
They were tuned using the cross-validation algorithm~\cite{Hastie}. The mean validation loss was used as the figure of
merit for the optimization over a five-fold splitting of the training sample. The network has been trained on a single NVIDIA GeForce GTX 1080 Ti GPU.

Random sampling was used to select 50 of 120 grid points in hyperparameter space, where the grid is defined by the following:
\begin{itemize}
\item dropout rate: $do \in  [0.1, 0.2, 0.3, 0.4]$,
\item learning rate: $lr \in [10^{-2}, 10^{-3}, 10^{-4},  10^{-5},  10^{-6}]$ and
\item number of hidden layers: varied between 3 and 8.
\end{itemize}

The number of nodes in the last 3 hidden layers of the DNN was set to [512, 256, 128] respectively, while the number of nodes of the remaining layers was set to 1024. 
A number of configurations were found to provide comparable performance. Of these, the network with the smallest number of trainable parameters was chosen. The parameters and their values are: $do = 0.1$, $lr = 0.001$, and 6 hidden layers with [1024, 1024, 1024, 512, 256, 128] nodes. This architecture has a total of about 2.8 million trainable parameters.

\subsection{Training set \texorpdfstring{\pt}{pt}  composition\label{sec:TrainingStatistics}}

The number of events as a function of the \PQb jet \pt spectrum in the training sample spans six orders of magnitude, as shown in Fig.~\ref{fig:bjets}~(\cmsLeft). This
means that, during the training, the DNN is exposed to many more jets with low \pt. In situations like this, one might expect worse performance for high-\pt jets. To check if this is an issue, emphasis was given to the high \pt part of the sample. About
95\% of the jets with \pt below 400 \GeV were removed in order to reproduce the same exponential shape observed above 400 \GeV.  We found that the DNN trained on this subsample of events showed no improvement for high \pt jets but did have up to 0.5\% degradation of the relative jet energy resolution. For this reason, the final DNN is trained on the full sample.

\section{Results\label{sec:Results}}

The performance of the \PQb jet regression was evaluated by comparing the \PQb jet energy resolution and
scale (defined as the most probable value of the $\pTjgen / \pTjreco$ distribution), before and after the energy correction, on a test sample that is statistically
independent from those used for training and validation. Different physics processes were included in the test set in
order to evaluate the performance of the algorithm on \PQb jets with different kinematics.
The processes employed in the test sample are:
\begin{itemize}
\item \ttbar: top quark-antiquark pair production (independent of the training data set),
\item \ZHbbll: associated production of a Higgs boson with a {\PZ} boson, where the {\PZ} boson decays to a pair of same-flavor, opposite-charge electrons or muons,
and the Higgs boson decays to $\bbbar$,
\item \HHbbgg: double Higgs boson produced via gluon fusion with one Higgs boson decaying to $\bbbar$, and the other to a pair of photons, assuming both standard model (SM) and beyond SM kinematics. In the latter case, the double Higgs signal originates from the decay of a spin-0 resonance
with a mass of 500 or 700\GeV.
\end{itemize}

\begin{figure*}[h!t]
\centering
\includegraphics[width=0.32\textwidth]{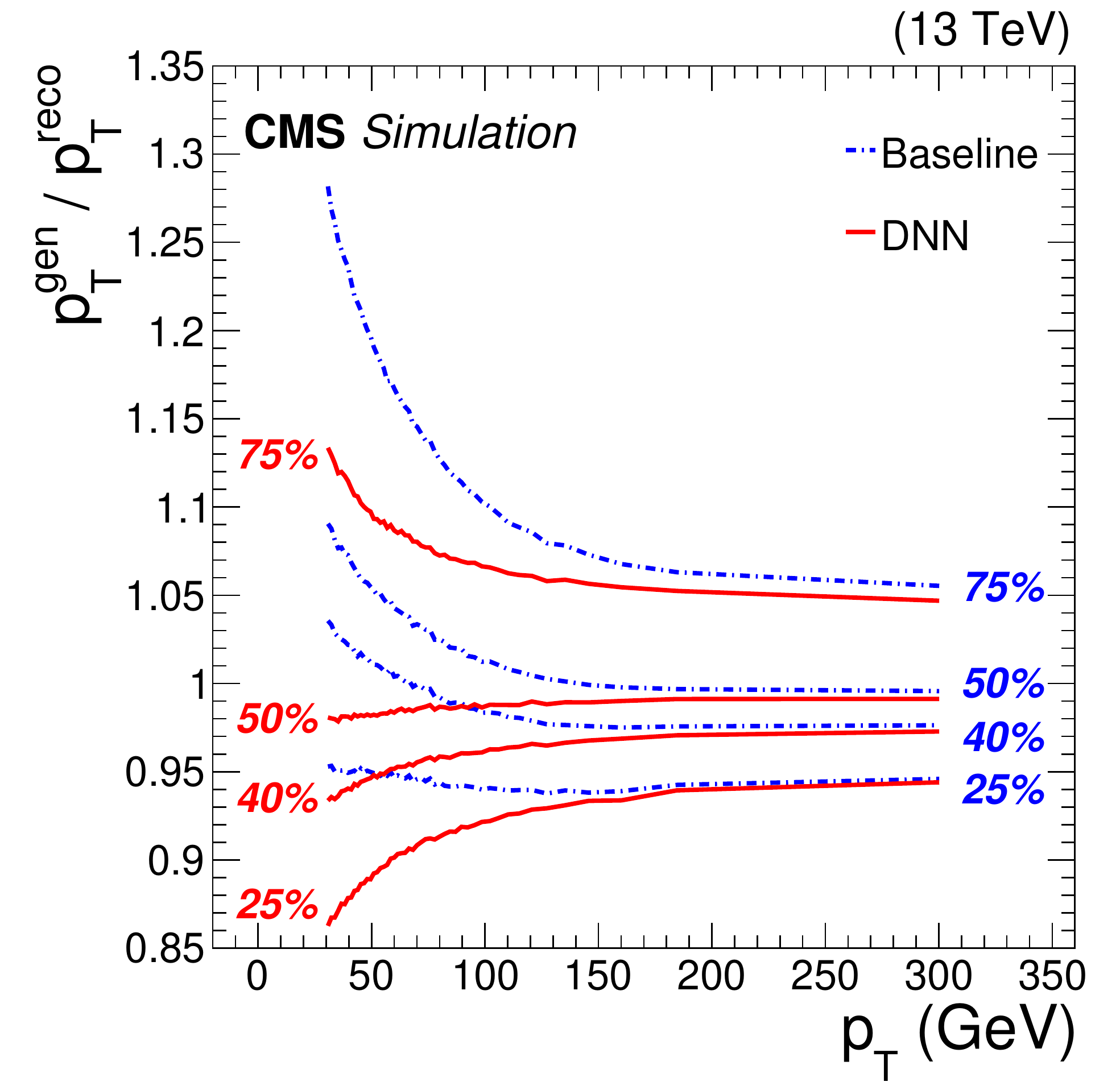}
\includegraphics[width=0.32\textwidth]{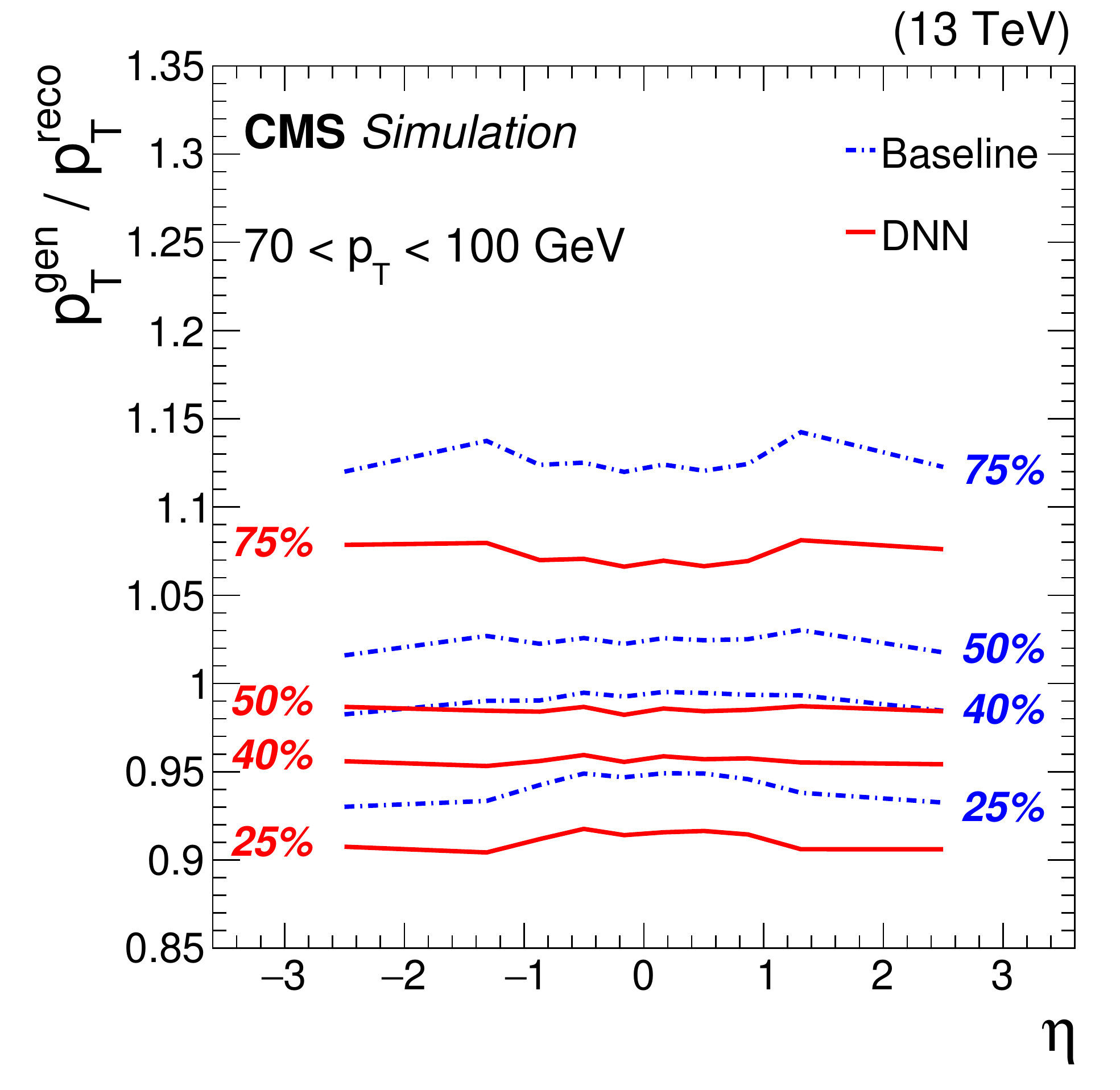}
\includegraphics[width=0.32\textwidth]{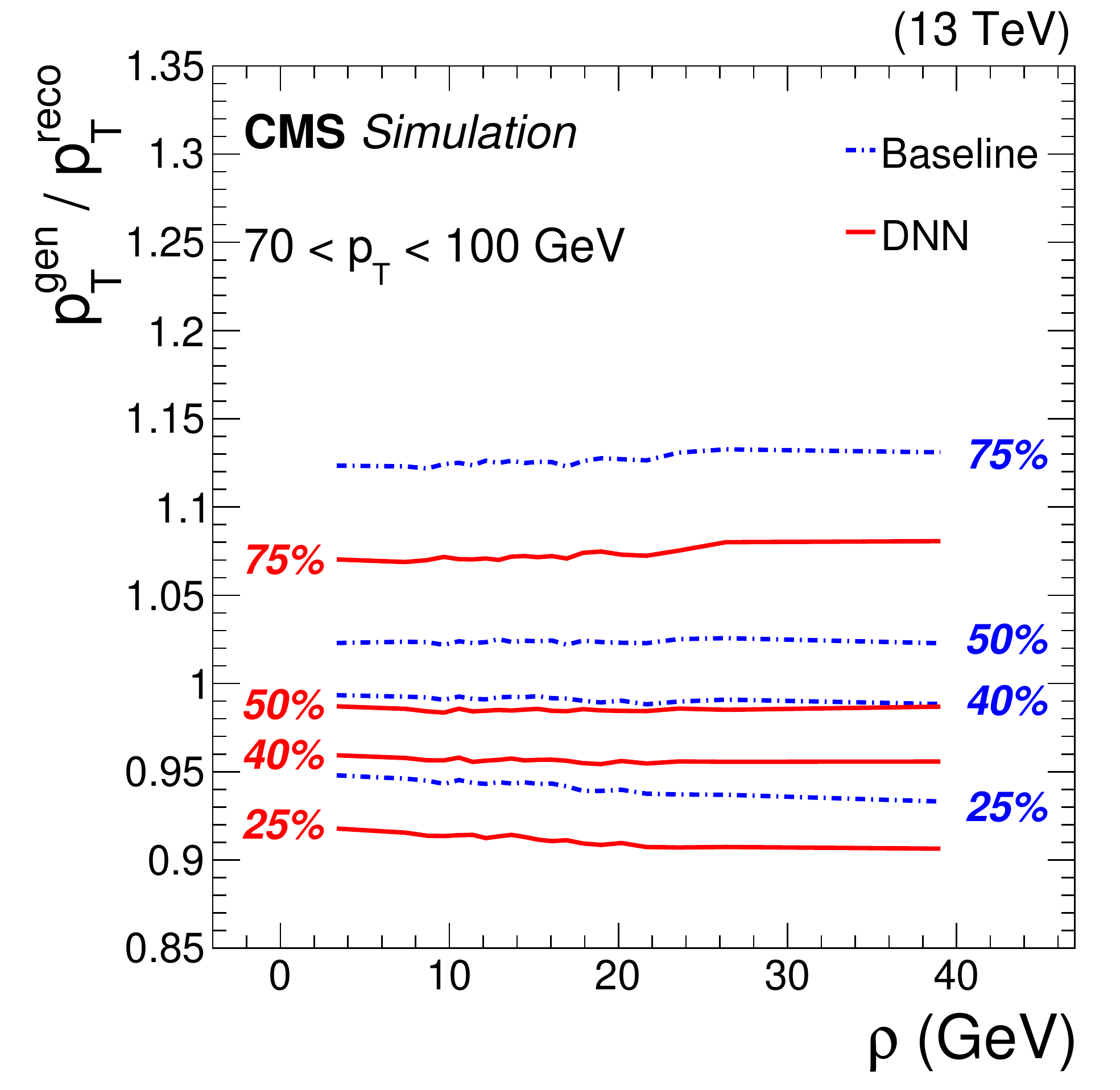}
\caption{  The 25, 40, 50, and 75\% quantiles are shown for the \PQb jet energy scale $\pTjgen / \pTjreco$ distribution before (blue dashdot) and after (red solid) applying the regression correction as a function of jet \pt (left), $\eta$ (center), and $\rho$ (right). The $\eta$ and $\rho$ distributions are shown for jets with
\pt $\in$ [70, 100]\GeV. \label{fig:scaleimprovement}}
\end{figure*}

Figure~\ref{fig:scaleimprovement} shows the 25, 40, 50, and 75\% quantiles of the target distribution before and after applying the DNN \PQb jet energy corrections, as a function of jet \pt, $\eta$, and $\rho$. The results are obtained for \PQb jets from the \ttbar test sample. The 40\% quantile has been found to be a good approximation of the most probable value of the target distribution. In addition, the 40\% quantile validates the performance on a quantile not used in the training.  It can be seen that after DNN corrections, the distribution becomes narrower, and its median and 40\% quantile exhibit smaller dependence on jet $\pt$, $\eta$, and the median event energy density $\rho$.

\begin{figure*}[h!t]
\centering
\includegraphics[width=0.32\textwidth]{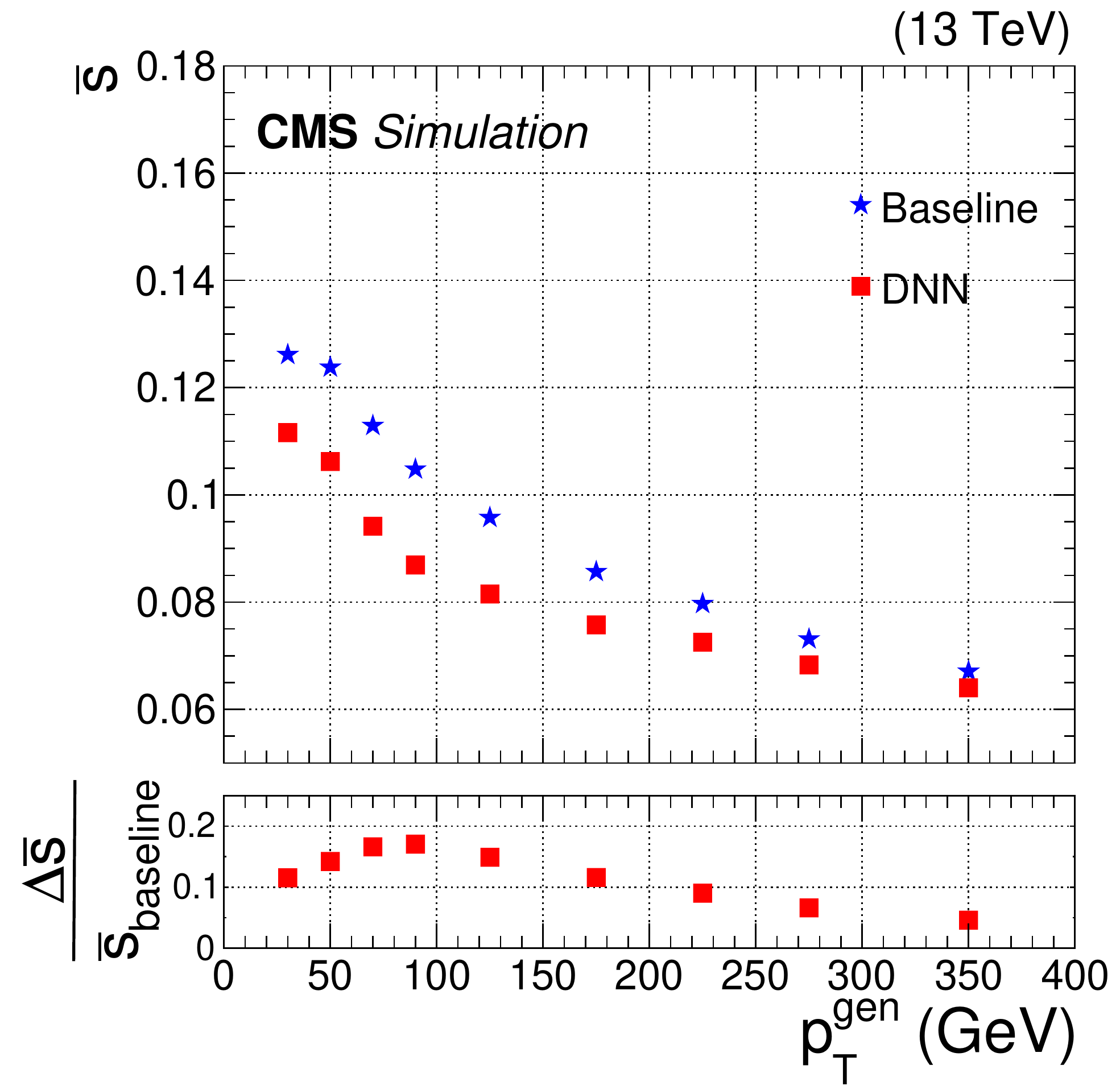}
\includegraphics[width=0.32\textwidth]{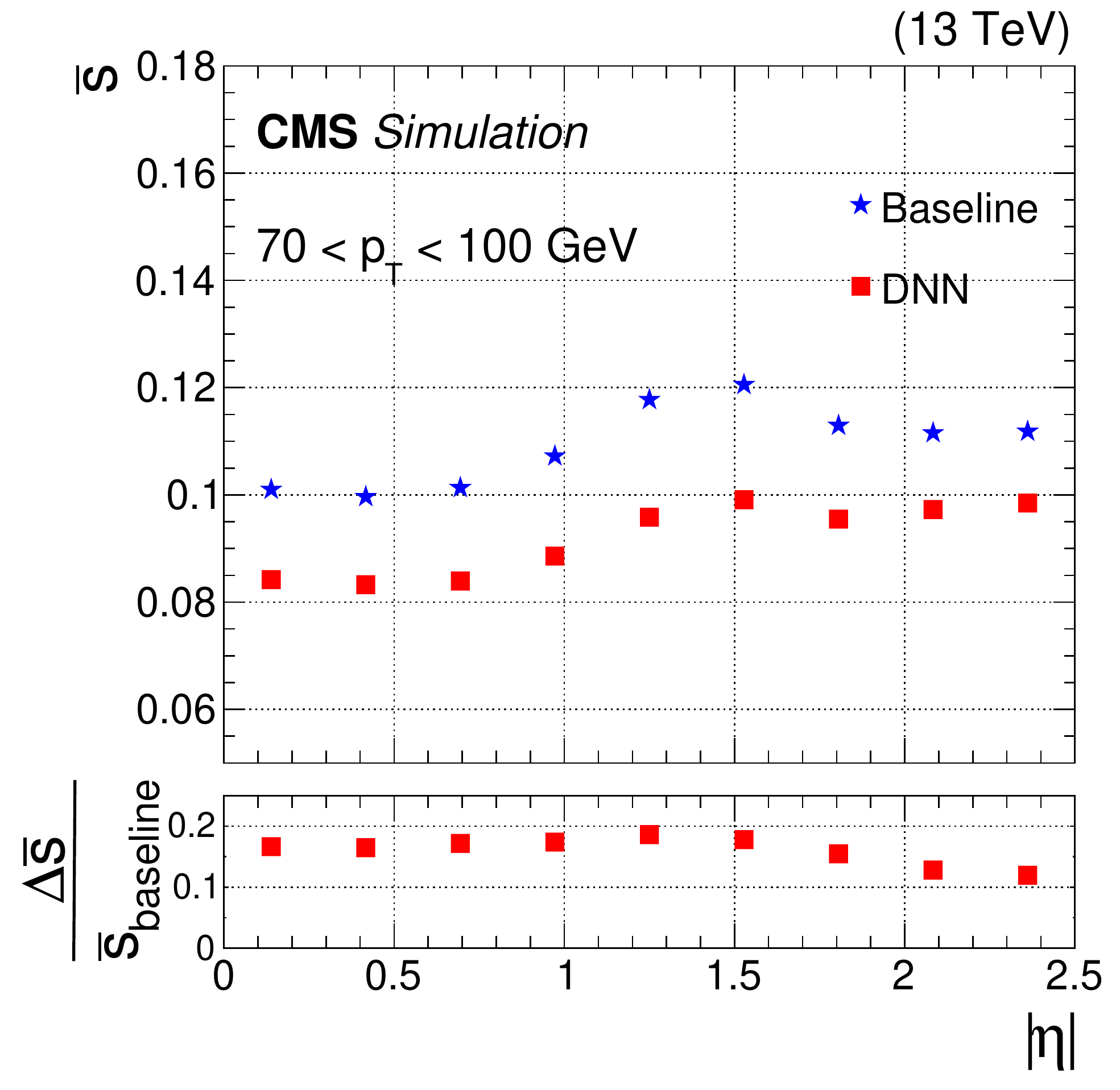}
\includegraphics[width=0.32\textwidth]{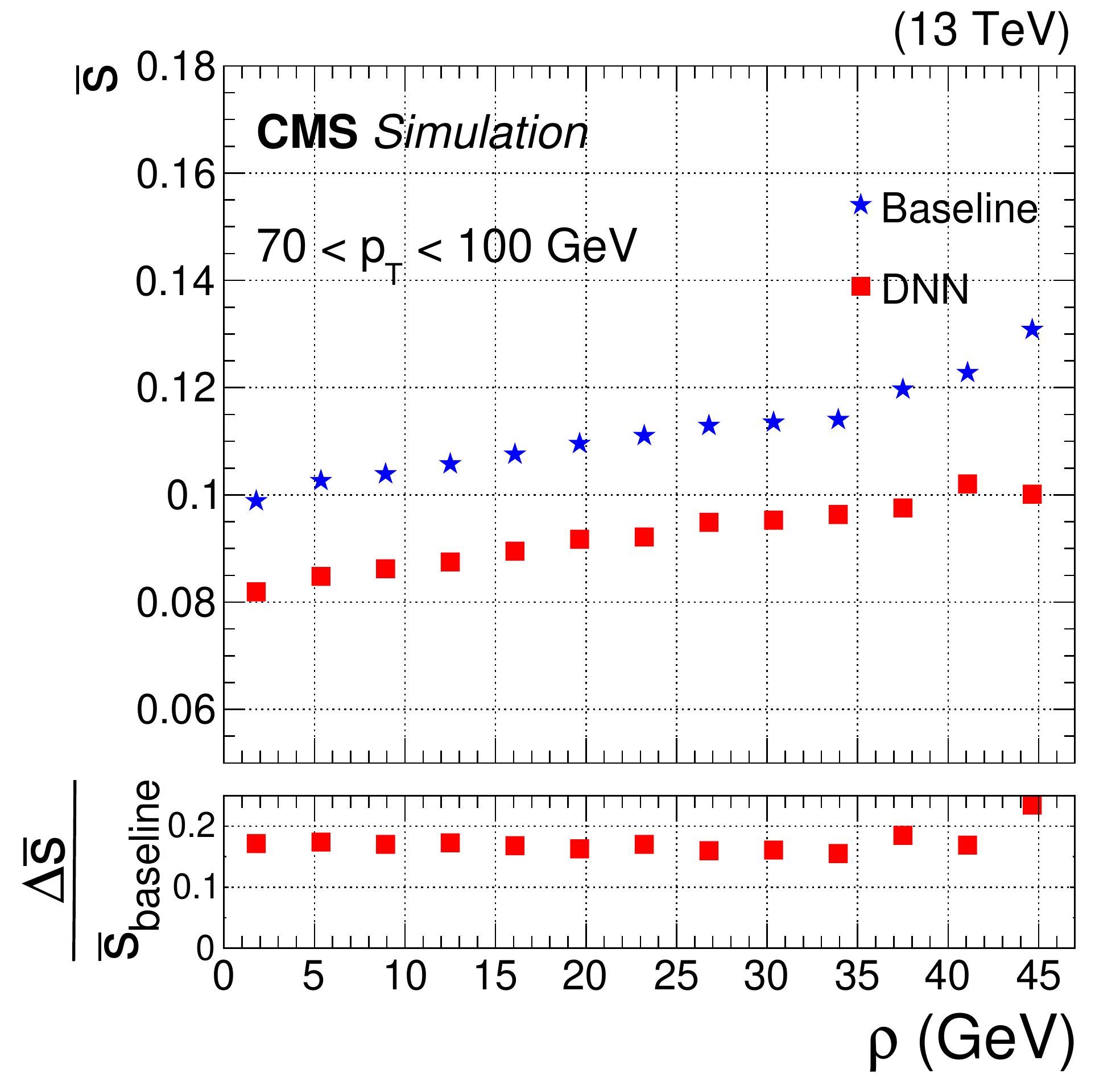}
\caption{Relative jet energy resolution, \sbar, as a function of generator-level jet $\pTjgen$ (left), $\eta$ (center), and $\rho$ (right) for \PQb jets from
  \ttbar MC events. The average \pt of these \PQb jets is 80\GeV. The $\eta$ and $\rho$ distributions are shown for jets with
\pt $\in$ [70, 100]\GeV. The blue stars and red squares represent
  \sbar  before and after the DNN correction, respectively. The relative difference $\Delta\sbar/\sbar_{\text{baseline}}$ between the \sbar values before and after DNN corrections is shown in the lower panels. \label{fig:jetres_rhoeta}}
\end{figure*}

The jet energy resolution, $\mathrm{s}$, is estimated as half the difference between the 75\% ($q_{75}$) and 25\% ($q_{25}$) quantiles of the target distribution. To quantify the resolution improvement, we compared the relative jet energy resolution, \sbar, defined as:
\begin{linenomath*}
\begin{equation}
\centering
\sbar \equiv \frac{\mathrm{s}}{q_{40}} = \frac{q_{75} - q_{25}}{2}\frac{1}{q_{40}},
\end{equation}
\end{linenomath*}
where the resolution $\mathrm{s}$ is divided by $q_{40}$, the most probable value estimated as the 40\% quantile of the target distribution. The relative improvement on \sbar for \PQb jets for various physics processes is between 12 and 15\%, as can be seen from Table~\ref{tab:improvement}. Figure~\ref{fig:jetres_rhoeta} shows the value of \sbar obtained for \PQb jets from
the \ttbar test sample as a function of the generator-level $\pTjgen$ (left),
 $\eta$ (center), and $\rho$ (right).
The lower panels in Fig.~\ref{fig:jetres_rhoeta} show the relative improvements resulting from the DNN energy correction.
The observed behavior agrees with the expectation that the regression correction should optimize the jet energy resolution, while the baseline corrections aim for a flat response as a function of the jet generator-level $\pTjgen$ and $\eta$. For all physics processes considered, the per-jet relative resolution improvement is around 12--18\% for $\pt <100\GeV$, falling to around 5--9\% for $\pt >200\GeV$. This improvement translates into an improvement in sensitivity of the analyses that make use of \PQb jets in the final state. The improvement in the \PQb jet energy resolution brought by the regression is similar for \PQb jets with and without associated leptons. This demonstrates that the algorithm is able to correct not only for the undetected neutrinos in semileptonic decays of \PQb hadrons, but also for effects that may only be present in hadronic decays. In addition, the regression was shown to improve the response of light jets by about 3\%.

\begin{table}[h]
    \caption{\small Relative differences $\Delta\sbar/\sbar_\text{baseline}$ between the
\sbar values obtained before and after applying the DNN energy correction for \PQb jets
produced in the different physics processes indicated. \label{tab:improvement}}
\centering
    \begin{tabular}{ ll }
    \hline
     MC sample & Improvement\\ \hline
     \ttbar & 12.2\% \\
      \ZHbbll &  12.8\%\\
      \HHbbgg SM & 13.1\% \\
      \HHbbgg resonant 500\GeV  & 14.5\%\\
       \HHbbgg resonant 700\GeV & 13.1\%\\
    \hline
    \end{tabular}
\end{table}

\begin{figure}[h!t]
\centering
\includegraphics[width=0.49\textwidth]{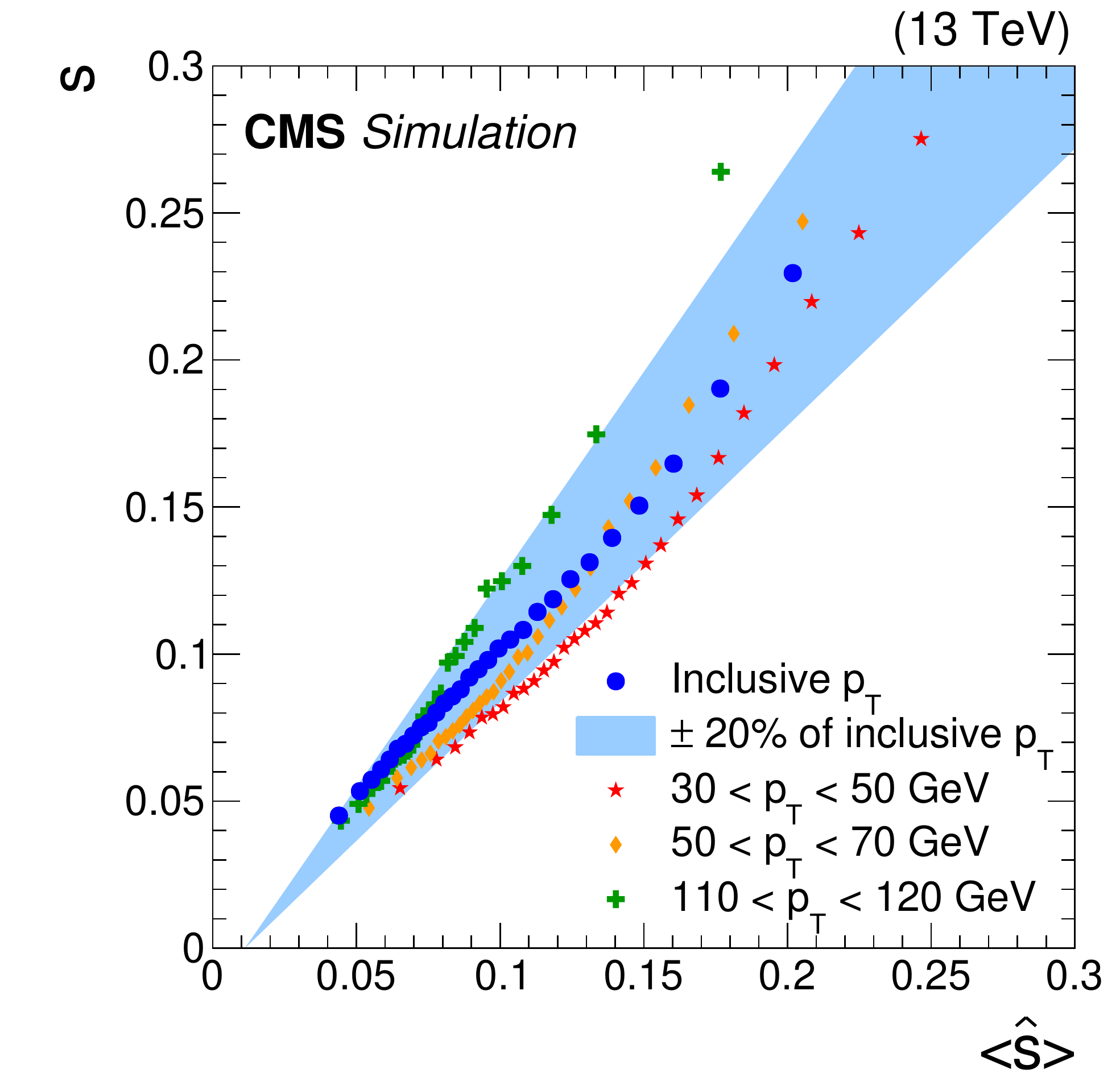}
\caption{ Correlation between jet energy resolution $\mathrm{s}$ and the average jet energy resolution estimator $\langle\hat{\mathrm{s}}\rangle$ for \PQb jets from \ttbar MC events.
 The blue circles correspond to the inclusive \pt spectrum, while the blue band
represents 20\% up and down variations of the fitted $\langle\hat{\mathrm{s}}\rangle$ trend for the inclusive \pt spectrum. The red stars correspond to jets with
\pt $\in$ [30, 50]\GeV, orange diamonds to \pt $\in$ [50, 70]\GeV, and green crosses to \pt $\in$ [110,120]\GeV.
 \label{fig:resolutionInclusive}}
\end{figure}

Knowledge of jet energy resolution on a jet-by-jet basis can be exploited in analyses searching for resonant production of \PQb jet pairs to increase their sensitivity. We have checked the correlation between the jet resolution $\mathrm{s}$ and the value of the per-jet resolution estimator, $\hat{\mathrm{s}}$, provided by the DNN:
\begin{linenomath*}
\begin{equation}
\centering
\hat{\mathrm{s}} \equiv \frac{1}{2}(\hat{y}_{75\%} - \hat{y}_{25\%}).
\end{equation}
\end{linenomath*}
To do this, the sample of \PQb jets was split into several equally populated bins in $\hat{\mathrm{s}}$. In each bin, the value of $\mathrm{s}$ is computed as half the difference between the $q_{75}$ and $q_{25}$ quantiles of the target distribution, and compared to the average resolution estimator $\langle\hat{\mathrm{s}}\rangle$.
Figure~\ref{fig:resolutionInclusive} shows the
correlation between $\mathrm{s}$ and the $\langle\hat{\mathrm{s}}\rangle$ values for the inclusive \pt spectrum and for several bins in \pt.
A linear dependence with slope near unity confirms that the per-jet energy resolution estimator $\hat{\mathrm{s}}$ correctly represents the jet resolution.
We observe that deviations of the slope from unity from the linear behavior are roughly compatible within 20\% of the $\hat{\mathrm{s}}$ value.

\begin{figure}[h!t]
\centering
\includegraphics[width=0.49\textwidth]{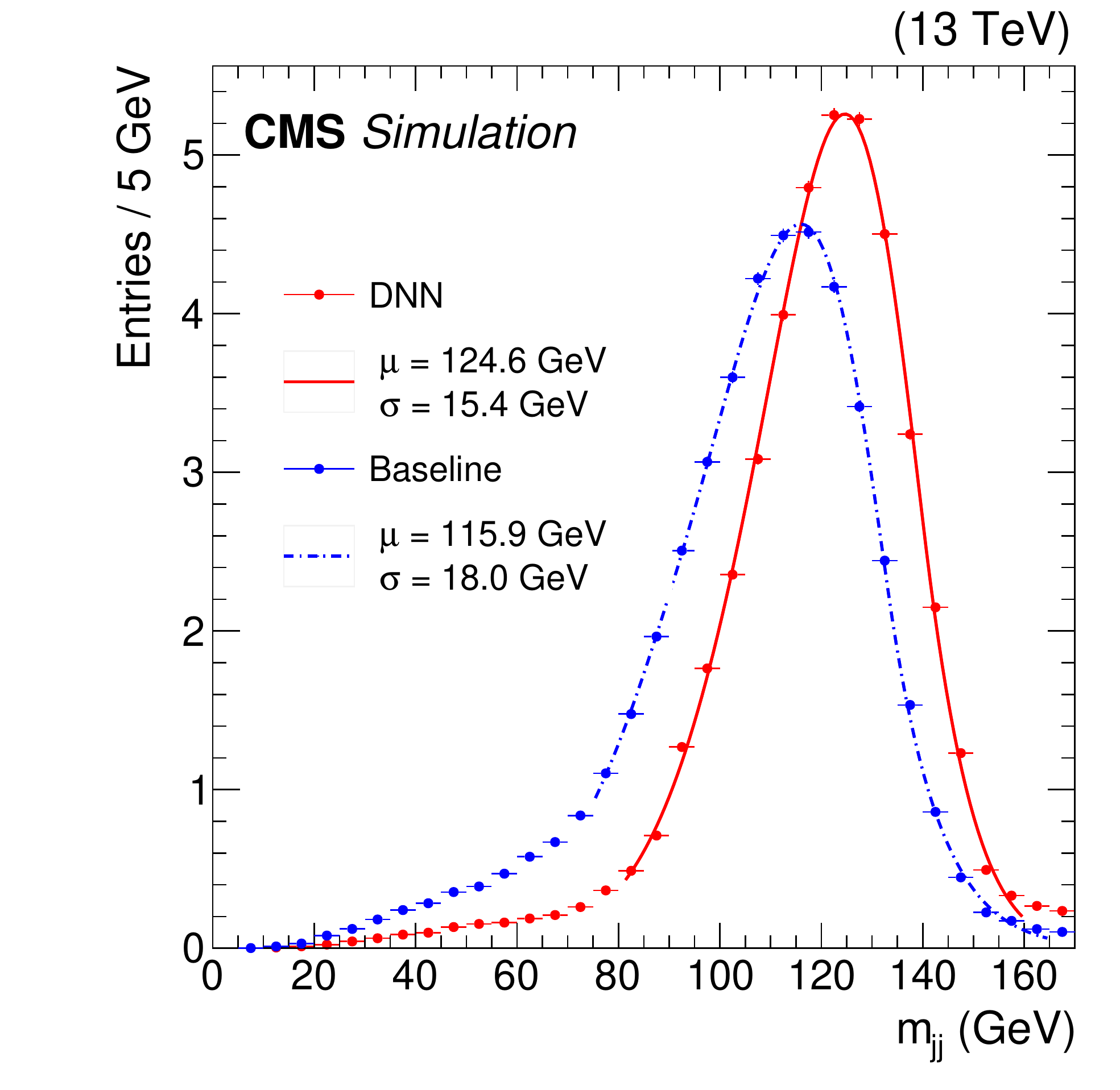}
\caption{\small Dijet invariant mass distributions for simulated samples of \ZHbbll events, where two jets and two leptons were selected. Distributions are shown before (dotted blue) and after (solid red) applying the \PQb jet energy corrections.
A Bukin function~\cite{Bukin} was used to fit the distribution. The fitted mean and width of the core of each distribution are displayed in the figure.
 \label{fig:dijetmassNadya}}
\end{figure}

While the improvements described above are quoted at the single jet level, many physics analyses use the invariant mass of the two \PQb jet system as a discriminating variable for signal extraction.
The improvement in the resolution of the dijet invariant mass is generally bigger than
that for a single jet because the energy corrections effectively equalize the energy scale of the two jets, while also improving the jet resolution.
 To estimate the dijet resolution improvement events with two leptons and two jets were selected from the \ZHbbll sample: jets were required to have $\pt$ larger than 20\GeV, absolute value of
$\eta$ below 2.4, and be compatible with the hadronisation of \PQb quarks, referred to as ``\PQb-tagged''~\cite{Sirunyan:2017ezt} jets in the following. The selection criteria for the \PQb-tagged jets correspond to a 70\% \PQb jet tagging efficiency with a 1\% misidentification rate for light-flavor or gluon jets.
 Leptons were required to have a \pt larger than 20\GeV, while the lepton pairs were required to be compatible with the decay of a {\PZ} boson, requiring
their invariant mass to be within 20\GeV of the mass of the {\PZ} boson.
The {\PZ} boson was required to have a transverse momentum larger than 150\GeV. An improvement of about 20\% in
the dijet invariant mass resolution in the \ZHbbll sample can be observed in
Fig.~\ref{fig:dijetmassNadya}. A Bukin function~\cite{Bukin} was used to fit the core of the distribution in
Fig.~\ref{fig:dijetmassNadya}. The fit is performed in the range [75, 165]\GeV for the baseline and [81,160]\GeV for the DNN corrected distribution.

In addition, a dedicated study was performed to test how well the algorithm performance
can be transferred from Monte Carlo simulations to the domain of {\Pp}{\Pp} collision data. A set of {\PZ} boson candidates decaying to a pair of charged leptons was extracted from {\Pp}{\Pp} collisions recorded by the CMS experiment in 2017.
A standard set of requirements~\cite{Chatrchyan:2011ds,Khachatryan:2014gga} was applied to select events with electron or muon pairs compatible with having originated from the decay of a {\PZ} boson.
Events were further required to have at least one \PQb-tagged jet. The jet with the largest $\pt$ was required to have $\abs{\eta} < 2$, while the \pt of the dilepton system was required to be larger than 100\GeV. The \pt balance between the {\PZ} boson and the \PQb-tagged jet candidate was enforced by requiring that extra jets have a \pt less than 30\% of the {\PZ} \pt to suppress events with additional hadronic activity.
Events satisfying these requirements were used to evaluate the agreement between data and MC simulations. In addition, the resolution of the jets was measured by extrapolating to zero additional hadronic activity following the methodology described in Ref.~\cite{Chatrchyan:2011ds}.

\begin{figure}[h!t]
\centering
\includegraphics[width=0.49\textwidth]{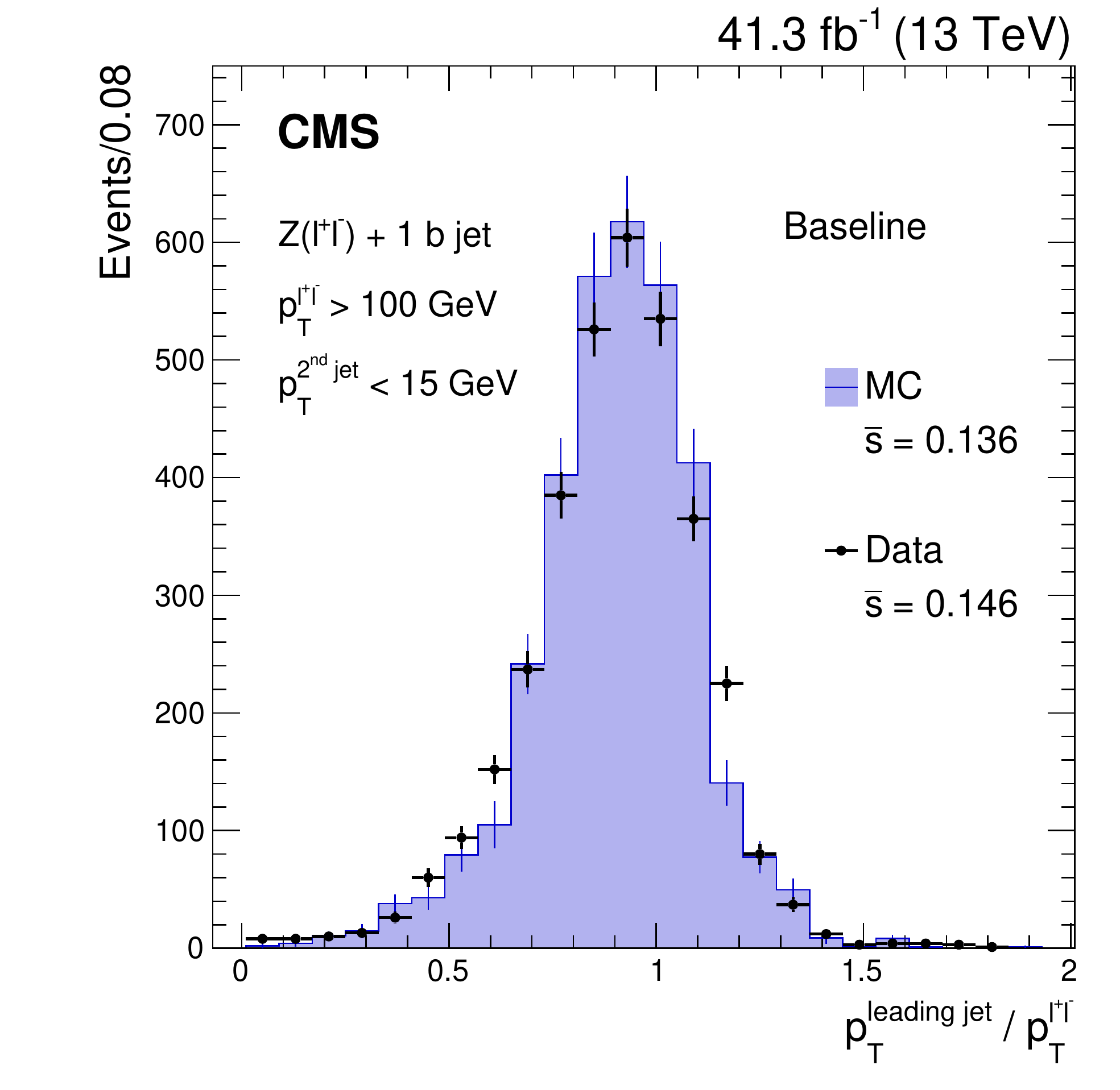}
\includegraphics[width=0.49\textwidth]{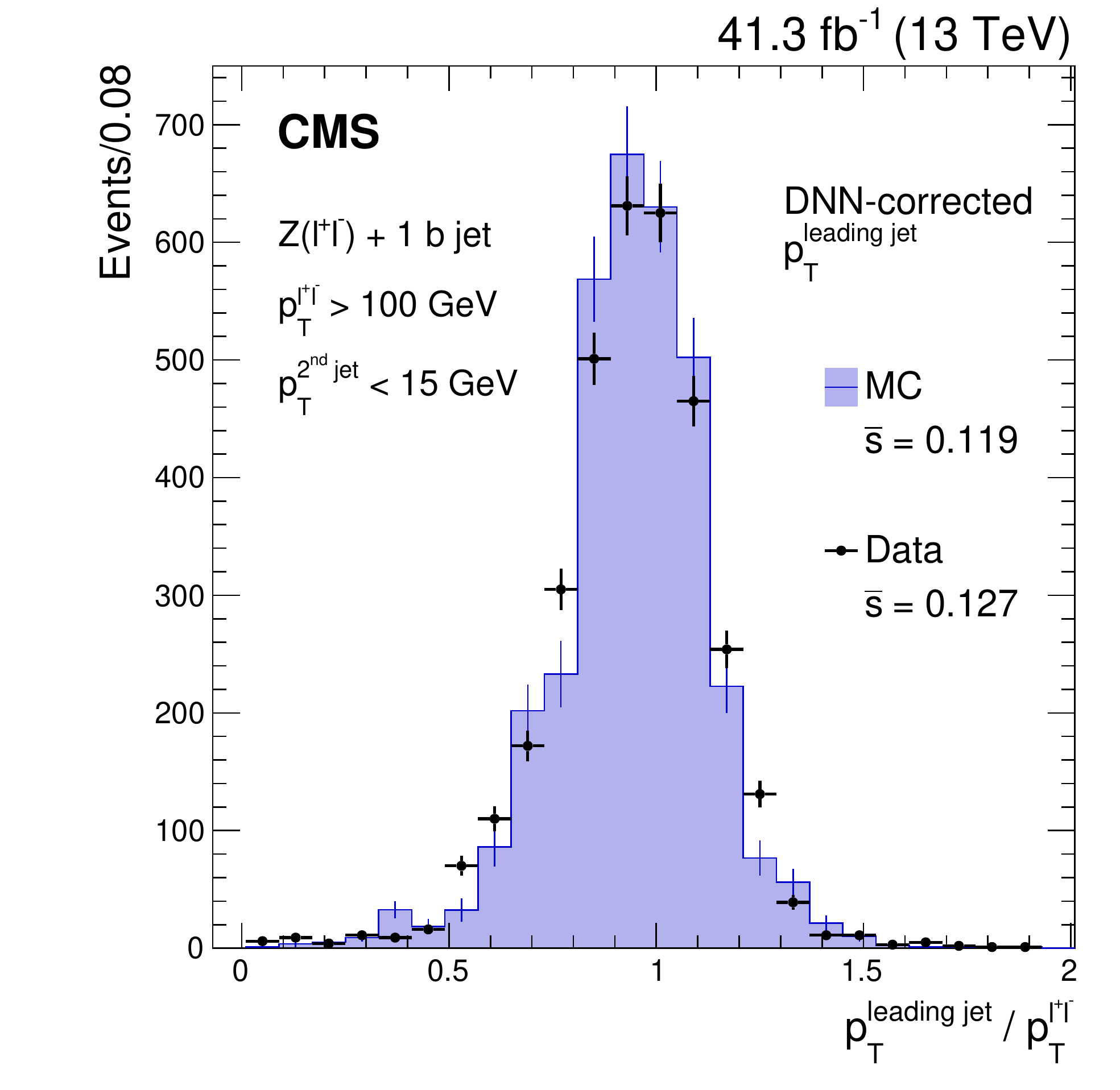}
\caption{ Distribution of the ratio between the transverse momentum of the leading \PQb-tagged jet and that of the dilepton system from the decay of the {\PZ} boson.
Distributions are shown before (\cmsLeft) and after (\cmsRight) applying the \PQb jet energy corrections. The \sbar values of the core distributions are included in the figures. The black points and histogram show the distributions for data and simulated events, respectively.
\label{fig:valid}}
\end{figure}

Figure~\ref{fig:valid} shows the ratio between the \pt of the leading jet and that of the dilepton system for events in which the \pt of the subleading jet is less than 15\GeV. The \cmsLeft and \cmsRight panels show the distributions obtained before and after applying the DNN-based corrections, respectively. It can be seen that the effect of the corrections is to reduce the width of the distribution. Using the method detailed in Ref.~\cite{Chatrchyan:2011ds}, the double ratio of the relative jet resolution \sbar measured in data and in simulated events was found to be $1.1 \pm 0.1$ before and after applying the DNN-based corrections. This validates that the resolution improvement achieved in simulated events is successfully transferred to the data domain.

\section{Summary}\label{sec:Conclusion}

We have described an algorithm that makes it possible to obtain point and dispersion estimates of the
energy of jets arising from \PQb quarks in proton-proton collisions.
We trained a deep, feed-forward neural network, with inputs based on jet composition and shape information, and on
properties of the associated reconstructed secondary vertex for a sample of simulated \PQb jets arising from the decays of top quark-antiquark pairs.
The neural network simultaneously finds robust mean, 25 and 75\% quantile estimators for the energy of a \PQb jet. The mean estimator is based on the Huber loss
function and is used as an energy correction, while the 25 and 75\% quantile estimators are used to build a jet-by-jet resolution estimator, defined as half the difference between these quantiles.

The DNN-based algorithm leverages the information contained in a large training data set consisting
of nearly 100 million simulated \PQb jets, and improves the resolution of the \PQb jet
energy by 12--15\% relative to that which is found after baseline corrections. An improvement of about 20\% is observed in the resolution of the invariant mass of \PQb jet pairs
resulting from the decay of a Higgs boson produced in association with a {\PZ} boson.
The resolution estimator is further shown to predict the resolution of \PQb jets with an accuracy of 20\% over a \pt range between 30 and 350\GeV.
Events containing a dilepton decay of a {\PZ} boson produced in association with a \PQb jet are used to validate the performance of the algorithm on proton-proton collision data recorded with the CMS detector.
The jet energy resolution improvement observed in data is consistent with that found in simulation.

The results described here are being used by the CMS Collaboration in several physics analyses targeting final states containing \PQb jets, including the observation of the Higgs boson decay to $\bbbar$~\cite{Sirunyan:2018kst}.

\begin{acknowledgments}
We congratulate our colleagues in the CERN accelerator departments for the excellent performance of the LHC and thank the technical and administrative staffs at CERN and at other CMS institutes for their contributions to the success of the CMS effort. In addition, we gratefully acknowledge the computing centers and personnel of the Worldwide LHC Computing Grid for delivering so effectively the computing infrastructure essential to our analyses. Finally, we acknowledge the enduring support for the construction and operation of the LHC and the CMS detector provided by the following funding agencies: BMBWF and FWF (Austria); FNRS and FWO (Belgium); CNPq, CAPES, FAPERJ, FAPERGS, and FAPESP (Brazil); MES (Bulgaria); CERN; CAS, MoST, and NSFC (China); COLCIENCIAS (Colombia); MSES and CSF (Croatia); RPF (Cyprus); SENESCYT (Ecuador); MoER, ERC IUT, PUT and ERDF (Estonia); Academy of Finland, MEC, and HIP (Finland); CEA and CNRS/IN2P3 (France); BMBF, DFG, and HGF (Germany); GSRT (Greece); NKFIA (Hungary); DAE and DST (India); IPM (Iran); SFI (Ireland); INFN (Italy); MSIP and NRF (Republic of Korea); MES (Latvia); LAS (Lithuania); MOE and UM (Malaysia); BUAP, CINVESTAV, CONACYT, LNS, SEP, and UASLP-FAI (Mexico); MOS (Montenegro); MBIE (New Zealand); PAEC (Pakistan); MSHE and NSC (Poland); FCT (Portugal); JINR (Dubna); MON, RosAtom, RAS, RFBR, and NRC KI (Russia); MESTD (Serbia); SEIDI, CPAN, PCTI, and FEDER (Spain); MOSTR (Sri Lanka); Swiss Funding Agencies (Switzerland); MST (Taipei); ThEPCenter, IPST, STAR, and NSTDA (Thailand); TUBITAK and TAEK (Turkey); NASU (Ukraine); STFC (United Kingdom); DOE and NSF (USA).

\hyphenation{Rachada-pisek} Individuals have received support from the Marie-Curie program and the European Research Council and Horizon 2020 Grant, contract Nos.\ 675440, 752730, and 765710 (European Union); the Leventis Foundation; the A.P.\ Sloan Foundation; the Alexander von Humboldt Foundation; the Belgian Federal Science Policy Office; the Fonds pour la Formation \`a la Recherche dans l'Industrie et dans l'Agriculture (FRIA-Belgium); the Agentschap voor Innovatie door Wetenschap en Technologie (IWT-Belgium); the F.R.S.-FNRS and FWO (Belgium) under the ``Excellence of Science -- EOS" -- be.h project n.\ 30820817; the Beijing Municipal Science \& Technology Commission, No. Z181100004218003; the Ministry of Education, Youth and Sports (MEYS) of the Czech Republic; the Deutsche Forschungsgemeinschaft (DFG) under Germany’s Excellence Strategy -- EXC 2121 ``Quantum Universe" -- 390833306; the Lend\"ulet (``Momentum") Program and the J\'anos Bolyai Research Scholarship of the Hungarian Academy of Sciences, the New National Excellence Program \'UNKP, the NKFIA research grants 123842, 123959, 124845, 124850, 125105, 128713, 128786, and 129058 (Hungary); the Council of Science and Industrial Research, India; the HOMING PLUS program of the Foundation for Polish Science, cofinanced from European Union, Regional Development Fund, the Mobility Plus program of the Ministry of Science and Higher Education, the National Science Center (Poland), contracts Harmonia 2014/14/M/ST2/00428, Opus 2014/13/B/ST2/02543, 2014/15/B/ST2/03998, and 2015/19/B/ST2/02861, Sonata-bis 2012/07/E/ST2/01406; the National Priorities Research Program by Qatar National Research Fund; the Ministry of Science and Education, grant no. 14.W03.31.0026 (Russia); the Programa Estatal de Fomento de la Investigaci{\'o}n Cient{\'i}fica y T{\'e}cnica de Excelencia Mar\'{\i}a de Maeztu, grant MDM-2015-0509 and the Programa Severo Ochoa del Principado de Asturias; the Thalis and Aristeia programs cofinanced by EU-ESF and the Greek NSRF; the Rachadapisek Sompot Fund for Postdoctoral Fellowship, Chulalongkorn University and the Chulalongkorn Academic into Its 2nd Century Project Advancement Project (Thailand); the Nvidia Corporation; the Welch Foundation, contract C-1845; and the Weston Havens Foundation (USA).
\end{acknowledgments}

\bibliography{auto_generated}

\providecommand{\href}[2]{#2}\begingroup\raggedright\begin{thebibliography}{10}%
\makeatletter
\providecommand{\hrefCMSnoop }[0]{\@secondoftwo}%
\makeatother
\providecommand{\doi}{\texttt{doi:}\begingroup \urlstyle{tt}\Url}

\bibitem{Aad:2012tfa}
\hrefCMSnoop {}{{ATLAS Collaboration}, ``{Observation of a new particle in the
  search for the standard model Higgs boson with the ATLAS detector at the
  LHC}'',} \textit{ Phys. Lett. B} \textbf{ 716} (2012) 1,
  \href{http://dx.doi.org/10.1016/j.physletb.2012.08.020}{\doi{10.1016/j.physletb.2012.08.020}},
\href{http://www.arXiv.org/abs/1207.7214}{\texttt{arXiv:1207.7214}}.

\bibitem{Chatrchyan:2012xdj}
\hrefCMSnoop {}{{CMS Collaboration}, ``{Observation of a new boson at a mass of
  125 GeV with the CMS experiment at the LHC}'',} \textit{ Phys. Lett. B}
  \textbf{ 716} (2012) 30,
  \href{http://dx.doi.org/10.1016/j.physletb.2012.08.021}{\doi{10.1016/j.physletb.2012.08.021}},
\href{http://www.arXiv.org/abs/1207.7235}{\texttt{arXiv:1207.7235}}.

\bibitem{Chatrchyan:1529911}
\hrefCMSnoop {}{{CMS Collaboration}, ``A new boson with a mass of 125 {GeV}
  observed with the {CMS} experiment at the {Large Hadron Collider}'',}
  \textit{ Science} \textbf{ 338} (2012) 1569,
  \href{http://dx.doi.org/10.1126/science.1230816}{\doi{10.1126/science.1230816}}.

\bibitem{Aad:2014eva}
\hrefCMSnoop {}{{ATLAS Collaboration}, ``{Measurements of Higgs boson
  production and couplings in the four-lepton channel in pp collisions at
  center-of-mass energies of 7 and 8 TeV with the ATLAS detector}'',} \textit{
  Phys. Rev. D} \textbf{ 91} (2015) 012006,
  \href{http://dx.doi.org/10.1103/PhysRevD.91.012006}{\doi{10.1103/PhysRevD.91.012006}},
\href{http://www.arXiv.org/abs/1408.5191}{\texttt{arXiv:1408.5191}}.

\bibitem{ATLAS:2014aga}
\hrefCMSnoop {}{{ATLAS Collaboration}, ``{Observation and measurement of Higgs
  boson decays to WW$^*$ with the ATLAS detector}'',} \textit{ Phys. Rev. D}
  \textbf{ 92} (2015) 012006,
  \href{http://dx.doi.org/10.1103/PhysRevD.92.012006}{\doi{10.1103/PhysRevD.92.012006}},
\href{http://www.arXiv.org/abs/1412.2641}{\texttt{arXiv:1412.2641}}.

\bibitem{Aad:2014eha}
\hrefCMSnoop {}{{ATLAS Collaboration}, ``{Measurement of Higgs boson production
  in the diphoton decay channel in pp collisions at center-of-mass energies of
  7 and 8 TeV with the ATLAS detector}'',} \textit{ Phys. Rev. D} \textbf{ 90}
  (2014) 112015,
  \href{http://dx.doi.org/10.1103/PhysRevD.90.112015}{\doi{10.1103/PhysRevD.90.112015}},
\href{http://www.arXiv.org/abs/1408.7084}{\texttt{arXiv:1408.7084}}.

\bibitem{Chatrchyan:2013mxa}
\hrefCMSnoop {}{{CMS Collaboration}, ``{Measurement of the properties of a
  Higgs boson in the four-lepton final state}'',} \textit{ Phys. Rev. D}
  \textbf{ 89} (2014) 092007,
  \href{http://dx.doi.org/10.1103/PhysRevD.89.092007}{\doi{10.1103/PhysRevD.89.092007}},
\href{http://www.arXiv.org/abs/1312.5353}{\texttt{arXiv:1312.5353}}.

\bibitem{Chatrchyan:2013iaa}
\hrefCMSnoop {}{{CMS Collaboration}, ``{Measurement of Higgs boson production
  and properties in the WW decay channel with leptonic final states}'',}
  \textit{ JHEP} \textbf{ 01} (2014) 096,
  \href{http://dx.doi.org/10.1007/JHEP01(2014)096}{\doi{10.1007/JHEP01(2014)096}},
\href{http://www.arXiv.org/abs/1312.1129}{\texttt{arXiv:1312.1129}}.

\bibitem{Khachatryan:2014ira}
\hrefCMSnoop {}{{CMS Collaboration}, ``{Observation of the diphoton decay of
  the Higgs boson and measurement of its properties}'',} \textit{ Eur. Phys. J.
  C} \textbf{ 74} (2014) 3076,
  \href{http://dx.doi.org/10.1140/epjc/s10052-014-3076-z}{\doi{10.1140/epjc/s10052-014-3076-z}},
\href{http://www.arXiv.org/abs/1407.0558}{\texttt{arXiv:1407.0558}}.

\bibitem{Sirunyan:2017khh}
\hrefCMSnoop {}{{CMS Collaboration}, ``{Observation of the Higgs boson decay to
  a pair of $\tau$ leptons with the CMS detector}'',} \textit{ Phys. Lett. B}
  \textbf{ 779} (2018) 283,
  \href{http://dx.doi.org/10.1016/j.physletb.2018.02.004}{\doi{10.1016/j.physletb.2018.02.004}},
\href{http://www.arXiv.org/abs/1708.00373}{\texttt{arXiv:1708.00373}}.

\bibitem{Aaboud:2018urx}
\hrefCMSnoop {}{{ATLAS Collaboration}, ``{Observation of Higgs boson production
  in association with a top quark pair at the LHC with the ATLAS detector}'',}
  \textit{ Phys. Lett. B} \textbf{ 784} (2018) 173,
  \href{http://dx.doi.org/10.1016/j.physletb.2018.07.035}{\doi{10.1016/j.physletb.2018.07.035}},
\href{http://www.arXiv.org/abs/1806.00425}{\texttt{arXiv:1806.00425}}.

\bibitem{Sirunyan:2018hoz}
\hrefCMSnoop {}{{CMS Collaboration}, ``{Observation of
  $\mathrm{t\overline{t}}$H production}'',} \textit{ Phys. Rev. Lett.} \textbf{
  120} (2018) 231801,
  \href{http://dx.doi.org/10.1103/PhysRevLett.120.231801}{\doi{10.1103/PhysRevLett.120.231801}},
\href{http://www.arXiv.org/abs/1804.02610}{\texttt{arXiv:1804.02610}}.

\bibitem{Sirunyan:2018kst}
\hrefCMSnoop {}{{CMS Collaboration}, ``{Observation of Higgs boson decay to
  bottom quarks}'',} \textit{ Phys. Rev. Lett.} \textbf{ 121} (2018) 121801,
  \href{http://dx.doi.org/10.1103/PhysRevLett.121.121801}{\doi{10.1103/PhysRevLett.121.121801}},
\href{http://www.arXiv.org/abs/1808.08242}{\texttt{arXiv:1808.08242}}.

\bibitem{Aaboud:2018zhk}
\hrefCMSnoop {}{{ATLAS Collaboration}, ``{Observation of $\mathrm{H \rightarrow
  \bbbar}$ decays and VH production with the ATLAS detector}'',} \textit{ Phys.
  Lett. B} \textbf{ 786} (2018) 59,
  \href{http://dx.doi.org/10.1016/j.physletb.2018.09.013}{\doi{10.1016/j.physletb.2018.09.013}},
\href{http://www.arXiv.org/abs/1808.08238}{\texttt{arXiv:1808.08238}}.

\bibitem{Aaltonen:2011bp0}
T.~Aaltonen\hrefCMSnoop {}{ {et~al.}, ``{Improved b jet energy correction for
  $\mathrm{H \rightarrow \bbbar}$ searches at CDF}'',}
  \href{http://www.arXiv.org/abs/1107.3026}{\texttt{arXiv:1107.3026}}.

\bibitem{Aaltonen:2011bp}
\hrefCMSnoop {}{{CDF} Collaboration, ``{Search for the standard model Higgs
  boson decaying to a $\bbbar$ pair in events with one charged lepton and large
  missing transverse energy using the full CDF data set}'',} \textit{ Phys.
  Rev. Lett.} \textbf{ 109} (2012) 111804,
  \href{http://dx.doi.org/10.1103/PhysRevLett.109.111804}{\doi{10.1103/PhysRevLett.109.111804}},
\href{http://www.arXiv.org/abs/1207.1703}{\texttt{arXiv:1207.1703}}.

\bibitem{Khachatryan:2015bnx}
\hrefCMSnoop {}{{CMS Collaboration}, ``{Search for the standard model Higgs
  boson produced through vector boson fusion and decaying to $\bbbar$}'',}
  \textit{ Phys. Rev. D} \textbf{ 92} (2015) 032008,
  \href{http://dx.doi.org/10.1103/PhysRevD.92.032008}{\doi{10.1103/PhysRevD.92.032008}},
\href{http://www.arXiv.org/abs/1506.01010}{\texttt{arXiv:1506.01010}}.

\bibitem{Huber:1964}
\hrefCMSnoop {}{P.~J. Huber, ``{Robust estimation of a location parameter}'',}
  \textit{ Ann. Math. Statist.} \textbf{ 35} (1994) 731,
  \href{http://dx.doi.org/10.1214/aoms/1177703732}{\doi{10.1214/aoms/1177703732}}.

\bibitem{RePEc}
\href
  {https://EconPapers.repec.org/RePEc:ecm:emetrp:v:46:y:1978:i:1:p:33-50}{R.~W.
  Koenker and G.~Bassett, ``Regression quantiles'',} \textit{ Econometrica}
  \textbf{ 46} (1978) 33.

\bibitem{Khachatryan:2015iwa}
\hrefCMSnoop {}{{CMS Collaboration}, ``{Performance of photon reconstruction
  and identification with the CMS Detector in proton-proton collisions at
  $\sqrt{s}$ = 8 TeV}'',} \textit{ JINST} \textbf{ 10} (2015) P08010,
  \href{http://dx.doi.org/10.1088/1748-0221/10/08/P08010}{\doi{10.1088/1748-0221/10/08/P08010}},
\href{http://www.arXiv.org/abs/1502.02702}{\texttt{arXiv:1502.02702}}.

\bibitem{Khachatryan:2015hwa}
\hrefCMSnoop {}{{CMS Collaboration}, ``{Performance of electron reconstruction
  and selection with the CMS detector in proton-proton collisions at $\sqrt{s}$
  = 8 TeV}'',} \textit{ JINST} \textbf{ 10} (2015) P06005,
  \href{http://dx.doi.org/10.1088/1748-0221/10/06/P06005}{\doi{10.1088/1748-0221/10/06/P06005}},
\href{http://www.arXiv.org/abs/1502.02701}{\texttt{arXiv:1502.02701}}.

\bibitem{Sirunyan:2019kia}
\hrefCMSnoop {}{{CMS Collaboration}, ``{Performance of missing transverse
  momentum reconstruction in proton-proton collisions at $\sqrt{s} =$ 13 TeV
  using the CMS detector}'',} \textit{ JINST} \textbf{ 14} (2019), no.~07,
  P07004,
  \href{http://dx.doi.org/10.1088/1748-0221/14/07/P07004}{\doi{10.1088/1748-0221/14/07/P07004}},
  \href{http://www.arXiv.org/abs/1903.06078}{\texttt{arXiv:1903.06078}}.

\bibitem{Chatrchyan:2008aa}
\hrefCMSnoop {}{{CMS Collaboration}, ``{The CMS experiment at the CERN LHC}'',}
  \textit{ JINST} \textbf{ 3} (2008) S08004,
\href{http://dx.doi.org/10.1088/1748-0221/3/08/S08004}{\doi{10.1088/1748-0221/3/08/S08004}}.

\bibitem{Sirunyan:2017ulk}
\hrefCMSnoop {}{{CMS Collaboration}, ``{Particle-flow reconstruction and global
  event description with the CMS detector}'',} \textit{ JINST} \textbf{ 12}
  (2017) P10003,
  \href{http://dx.doi.org/10.1088/1748-0221/12/10/P10003}{\doi{10.1088/1748-0221/12/10/P10003}},
\href{http://www.arXiv.org/abs/1706.04965}{\texttt{arXiv:1706.04965}}.

\bibitem{Cacciari:2008gp}
\hrefCMSnoop {}{M.~Cacciari, G.~P. Salam, and G.~Soyez, ``{The anti-$\kt$ jet
  clustering algorithm}'',} \textit{ JHEP} \textbf{ 04} (2008) 063,
  \href{http://dx.doi.org/10.1088/1126-6708/2008/04/063}{\doi{10.1088/1126-6708/2008/04/063}},
\href{http://www.arXiv.org/abs/0802.1189}{\texttt{arXiv:0802.1189}}.

\bibitem{Cacciari:2011ma}
\hrefCMSnoop {}{M.~Cacciari, G.~P. Salam, and G.~Soyez, ``{FastJet user
  Manual}'',} \textit{ Eur. Phys. J. C} \textbf{ 72} (2012) 1896,
  \href{http://dx.doi.org/10.1140/epjc/s10052-012-1896-2}{\doi{10.1140/epjc/s10052-012-1896-2}},
\href{http://www.arXiv.org/abs/1111.6097}{\texttt{arXiv:1111.6097}}.

\bibitem{Khachatryan_2017}
\hrefCMSnoop {}{{CMS Collaboration}, ``Jet energy scale and resolution in the
  {CMS} experiment in pp collisions at 8 {TeV}'',} \textit{ JINST} \textbf{ 12}
  (2017) P02014,
  \href{http://dx.doi.org/10.1088/1748-0221/12/02/P02014}{\doi{10.1088/1748-0221/12/02/P02014}},
  \href{http://www.arXiv.org/abs/1607.03663}{\texttt{arXiv:1607.03663}}.

\bibitem{Chatrchyan:2011ds}
\hrefCMSnoop {}{{CMS Collaboration}, ``{Determination of jet energy calibration
  and transverse momentum resolution in CMS}'',} \textit{ JINST} \textbf{ 6}
  (2011) P11002,
  \href{http://dx.doi.org/10.1088/1748-0221/6/11/P11002}{\doi{10.1088/1748-0221/6/11/P11002}},
\href{http://www.arXiv.org/abs/1107.4277}{\texttt{arXiv:1107.4277}}.

\bibitem{Campbell:2014kua}
\hrefCMSnoop {}{J.~M. Campbell, R.~K. Ellis, P.~Nason, and E.~Re, ``{Top-Pair
  production and decay at NLO matched with parton showers}'',} \textit{ JHEP}
  \textbf{ 04} (2015) 114,
  \href{http://dx.doi.org/10.1007/JHEP04(2015)114}{\doi{10.1007/JHEP04(2015)114}},
\href{http://www.arXiv.org/abs/1412.1828}{\texttt{arXiv:1412.1828}}.

\bibitem{Alwall:2014hca}
J.~Alwall\hrefCMSnoop {}{ {et~al.}, ``{The automated computation of tree-level
  and next-to-leading order differential cross sections, and their matching to
  parton shower simulations}'',} \textit{ JHEP} \textbf{ 07} (2014) 079,
  \href{http://dx.doi.org/10.1007/JHEP07(2014)079}{\doi{10.1007/JHEP07(2014)079}},
\href{http://www.arXiv.org/abs/1405.0301}{\texttt{arXiv:1405.0301}}.

\bibitem{Khachatryan_trig}
\hrefCMSnoop {}{{CMS Collaboration}, ``The {CMS} trigger system'',} \textit{
  JINST} \textbf{ 12} (2017) P01020,
  \href{http://dx.doi.org/10.1088/1748-0221/12/01/P01020}{\doi{10.1088/1748-0221/12/01/P01020}},
  \href{http://www.arXiv.org/abs/1609.02366}{\texttt{arXiv:1609.02366}}.

\bibitem{Sjostrand:2014zea}
T.~Sj{\"o}strand\hrefCMSnoop {}{ {et~al.}, ``An introduction to {PYTHIA}
  8.2'',} \textit{ Comput. Phys. Commun.} \textbf{ 191} (2015) 159,
  \href{http://dx.doi.org/10.1016/j.cpc.2015.01.024}{\doi{10.1016/j.cpc.2015.01.024}},
\href{http://www.arXiv.org/abs/1410.3012}{\texttt{arXiv:1410.3012}}.

\bibitem{Khachatryan:2015pea}
\hrefCMSnoop {}{{CMS Collaboration}, ``{Event generator tunes obtained from
  underlying event and multiparton scattering measurements}'',} \textit{ Eur.
  Phys. J. C} \textbf{ 76} (2016) 155,
  \href{http://dx.doi.org/10.1140/epjc/s10052-016-3988-x}{\doi{10.1140/epjc/s10052-016-3988-x}},
\href{http://www.arXiv.org/abs/1512.00815}{\texttt{arXiv:1512.00815}}.

\bibitem{Agostinelli:2002hh}
\hrefCMSnoop {}{{GEANT4} Collaboration, ``{GEANT4}---a simulation toolkit'',}
  \textit{ Nucl. Instrum. Meth. A} \textbf{ 506} (2003) 250,
\href{http://dx.doi.org/10.1016/S0168-9002(03)01368-8}{\doi{10.1016/S0168-9002(03)01368-8}}.

\bibitem{Cacciari:2007fd}
\hrefCMSnoop {}{M.~Cacciari and G.~P. Salam, ``{Pileup subtraction using jet
  areas}'',} \textit{ Phys. Lett. B} \textbf{ 659} (2008) 119,
  \href{http://dx.doi.org/10.1016/j.physletb.2007.09.077}{\doi{10.1016/j.physletb.2007.09.077}},
\href{http://www.arXiv.org/abs/0707.1378}{\texttt{arXiv:0707.1378}}.

\bibitem{Chatrchyan:1704291}
\hrefCMSnoop {}{{CMS Collaboration}, ``{Description and performance of track
  and primary-vertex reconstruction with the CMS tracker}'',} \textit{ JINST}
  \textbf{ 9} (2014) P10009,
  \href{http://dx.doi.org/10.1088/1748-0221/9/10/P10009}{\doi{10.1088/1748-0221/9/10/P10009}},
  \href{http://www.arXiv.org/abs/1405.6569}{\texttt{arXiv:1405.6569}}.

\bibitem{Sirunyan:2017ezt}
\hrefCMSnoop {}{{CMS Collaboration}, ``{Identification of heavy-flavour jets
  with the CMS detector in pp collisions at 13 TeV}'',} \textit{ JINST}
  \textbf{ 13} (2018) P05011,
  \href{http://dx.doi.org/10.1088/1748-0221/13/05/P05011}{\doi{10.1088/1748-0221/13/05/P05011}},
\href{http://www.arXiv.org/abs/1712.07158}{\texttt{arXiv:1712.07158}}.

\bibitem{Ioffe:2015ovl}
\href {http://proceedings.mlr.press/v37/ioffe15.html}{S.~Ioffe and C.~Szegedy,
  ``Batch normalization: accelerating deep network training by reducing
  internal covariate shift'',} in \textit{ Proceedings of Machine Learning
  Research}, volume~37, p.~448.
\newblock 2015.
\newblock
  \href{http://www.arXiv.org/abs/1502.03167}{\texttt{arXiv:1502.03167}}.

\bibitem{Maas:2013}
\href
  {http://ai.stanford.edu/~amaas/papers/relu_hybrid_icml2013_final.pdf}{A.~L.
  Maas {et~al.}, ``Rectifier nonlinearities improve neural network acoustic
  models'',} (2013).

\bibitem{chollet2015keras}
\href {https://keras.io}{F.~Chollet {et~al.}, ``Keras'',} 2015.
\newblock Software available from keras.io. \url {https://keras.io}.

\bibitem{tensorflow2015-whitepaper}
\href {http://tensorflow.org/}{M.~Abadi {et~al.}, ``{TensorFlow}: Large-scale
  machine learning on heterogeneous systems'',} 2015.
\newblock Software available from tensorflow.org. \url
  {http://tensorflow.org/}.

\bibitem{Adam}
\href {http://arxiv.org/abs/1412.6980}{D.~P. Kingma and J.~Ba, ``Adam: {A}
  method for stochastic optimization'',} (2014).
  \href{http://www.arXiv.org/abs/1412.6980}{\texttt{arXiv:1412.6980}}.

\bibitem{Hastie}
T.~Hastie, R.~Tibshirani, and J.~Friedman, ``The Elements of Statistical
  Learning''.
\newblock Springer-Verlag New York, 2nd edition, 2009.
\newblock
  \href{http://dx.doi.org/10.1007/978-0-387-84858-7}{\doi{10.1007/978-0-387-84858-7}}.

\bibitem{Bukin}
\hrefCMSnoop {}{A.~D. {Bukin}, ``Fitting function for asymmetric peaks'',}
  (2007). \href{http://www.arXiv.org/abs/0711.4449}{\texttt{arXiv:0711.4449}}.

\bibitem{Khachatryan:2014gga}
\hrefCMSnoop {}{{CMS Collaboration}, ``{Performance of the CMS missing
  transverse momentum reconstruction in pp data at $\sqrt{s}$ = 8 TeV}'',}
  \textit{ JINST} \textbf{ 10} (2015) P02006,
  \href{http://dx.doi.org/10.1088/1748-0221/10/02/P02006}{\doi{10.1088/1748-0221/10/02/P02006}},
\href{http://www.arXiv.org/abs/1411.0511}{\texttt{arXiv:1411.0511}}.

\end{thebibliography}\endgroup

\cleardoublepage \appendix\section{The CMS Collaboration \label{app:collab}}\begin{sloppypar}\hyphenpenalty=5000\widowpenalty=500\clubpenalty=5000\vskip\cmsinstskip
\textbf{Yerevan Physics Institute, Yerevan, Armenia}\\*[0pt]
A.M.~Sirunyan$^{\textrm{\dag}}$, A.~Tumasyan
\vskip\cmsinstskip
\textbf{Institut f\"{u}r Hochenergiephysik, Wien, Austria}\\*[0pt]
W.~Adam, F.~Ambrogi, T.~Bergauer, M.~Dragicevic, J.~Er\"{o}, A.~Escalante~Del~Valle, M.~Flechl, R.~Fr\"{u}hwirth\cmsAuthorMark{1}, M.~Jeitler\cmsAuthorMark{1}, N.~Krammer, I.~Kr\"{a}tschmer, D.~Liko, T.~Madlener, I.~Mikulec, N.~Rad, J.~Schieck\cmsAuthorMark{1}, R.~Sch\"{o}fbeck, M.~Spanring, D.~Spitzbart, W.~Waltenberger, C.-E.~Wulz\cmsAuthorMark{1}, M.~Zarucki
% ...
\vskip\cmsinstskip
\dag: Deceased\\
1:  Also at Vienna University of Technology, Vienna, Austria\\
% ...
\end{sloppypar}
\end{document}